\documentclass[12pt]{article}
\usepackage{graphicx,amsmath,amssymb,fullpage,times}
\usepackage[usenames]{color}
\usepackage[normalem]{ulem}
\usepackage{hyperref}
\usepackage[authoryear,colon,longnamesfirst,sectionbib]{natbib}
\usepackage{setspace}
\usepackage{threeparttable}
\usepackage[compact]{titlesec}

\makeatletter
\newcommand{\rmnum}[1]{\romannumeral #1}
\newcommand{\Rmnum}[1]{\expandafter\@slowromancap\romannumeral #1@}
\makeatother
\usepackage[footnotesize,bf]{caption}
\newcommand{\be}{\begin{enumerate}}
\newcommand{\ee}{\end{enumerate}}
\newcommand{\bi}{\begin{itemize}}
\newcommand{\ei}{\end{itemize}}
\newcommand{\bc}{\begin{center}}
\newcommand{\ec}{\end{center}}
\newcommand{\beq}{\begin{equation}}
\newcommand{\eeq}{\end{equation}}

\definecolor{gray}{gray}{.5}
\definecolor{litegray}{gray}{.8}
\usepackage[hang,flushmargin]{footmisc}
\usepackage[notablist, nofiglist, tablesfirst]{endfloat}
\newcounter{alphasect}

 \renewcommand\thesection{%
 \ifnum\value{alphasect}=1%
A
 \else
\ifnum\value{alphasect}=2%
B
\else
\ifnum\value{alphasect}=3%
C
\else
\ifnum\value{alphasect}=4%
D
\else
 \arabic{section}
 \fi\fi\fi\fi}%

 \newenvironment{asection}{%
 \setcounter{alphasect}{1}
 }{%
 \setcounter{alphasect}{0}
 }%

\usepackage{titlesec}
\titleformat{\section}
  {\normalfont\Large\bfseries\centering}{\thesection.}{1em}{}
\bibpunct[; ]{(}{)}{;}{a}{,}{;}

\begin{document}
\setlength{\baselineskip}{24pt} 
\thispagestyle{empty}

\title{\Large \onehalfspacing THERAPEUTIC HYPOTHERMIA: QUANTIFICATION OF THE TRANSITION OF CORE BODY TEMPERATURE USING THE FLEXIBLE MIXTURE BENT-CABLE MODEL FOR LONGITUDINAL DATA$^\S$\vspace{0.5cm}}
\author{\sc Shahedul A. Khan$^{1, \ast}$, Grace S. Chiu$^{2}$, and Joel A. Dubin$^{3}$}

\let\oldthefootnote\thefootnote
\renewcommand{\thefootnote}{\fnsymbol{footnote}}
\footnotetext[4]{On 2013-04-14, a revised version of this manuscript was accepted for publication in the {\it Australian and New Zealand Journal of Statistics}.}
\let\thefootnote\oldthefootnote

\let\oldthefootnote\thefootnote
\renewcommand{\thefootnote}{\fnsymbol{footnote}}
\footnotetext[1]{ Author to whom correspondence should be addressed}
\let\thefootnote\oldthefootnote

\let\oldthefootnote\thefootnote
\renewcommand{\thefootnote}{\arabic{footnote}}
\footnotetext[1]{Department of Mathematics and Statistics, University of Saskatchewan, Saskatoon, SK S7N 5E6, Canada.\\ email: s.khan@usask.ca}
\let\thefootnote\oldthefootnote

\let\oldthefootnote\thefootnote
\renewcommand{\thefootnote}{\arabic{footnote}}
\footnotetext[2]{CSIRO Mathematics, Informatics and Statistics, GPO Box 664, Canberra, ACT 2601, Australia.\\ email: grace.chiu@csiro.au}
\let\thefootnote\oldthefootnote

\let\oldthefootnote\thefootnote
\renewcommand{\thefootnote}{\arabic{footnote}}
\footnotetext[3]{Department of Statistics and Actuarial Science, School of Public Health and Health Systems, University of Waterloo, Waterloo, Ontario, N2L 3G1, Canada.\\ email: jdubin@uwaterloo.ca}
\let\thefootnote\oldthefootnote

\let\oldthefootnote\thefootnote
\renewcommand{\thefootnote}{\relax}
\footnotetext{\emph{Acknowledgments.} This work was partially supported by NSERC through Discovery Grants to G.~S.~Chiu (RGPIN 261806-05) and J.~A.~Dubin (RGPIN 327093-06), and by the Government of Ontario through Ontario Graduate Scholarships to S.~A.~Khan (000113006 \& 00012613. The authors thank Dr.~P.~S.~Reynolds, Department of Emergency Medicine and Virginia Commonwealth University Reanimation Science Center, Virginia Commonwealth University Medical Center, Richmond, VA USA, for permitting the rat data to be used in this article.}
\let\thefootnote\oldthefootnote

\date{}
\maketitle \vspace{-2cm}

\newpage
\begin{center}
\textbf{\Large Summary}\vspace{0.15cm}\\
\end{center}
By reducing core body temperature, $T_c$, induced hypothermia is a therapeutic tool to prevent brain damage resulting from physical trauma. However, all physiological systems begin to slow down due to hypothermia that in turn can result in increased risk of mortality. Therefore, quantification of the transition of $T_c$ to early hypothermia is of great clinical interest. Conceptually, $T_c$ may exhibit an either gradual or abrupt transition. Bent-cable regression is an appealing statistical tool to model such data due to the model's flexibility and greatly interpretable regression coefficients. It handles more flexibly models that traditionally have been handled by low-order polynomial models (for gradual transition) or piecewise linear changepoint models (for abrupt change). We consider a rat model for humans to quantify the temporal trend of $T_c$ to primarily address the question: What is the critical time point associated with a breakdown in the compensatory mechanisms following the start of hypothermia therapy? To this end, we develop a Bayesian modelling framework for bent-cable regression of longitudinal data to simultaneously account for gradual and abrupt transitions. Our analysis reveals that: (a) about 39\% of rats exhibit a gradual transition in $T_c$; (b) the critical time point is approximately the same regardless of transition type; (c) both transition types show a significant increase of $T_c$ followed by~a~significant~decrease. \vspace{0.2cm}

\newpage

\section{Introduction}
\label{s:introduction}
Longitudinal data arise in many areas of study, where measurements taken over time are nested within observational units drawn from some population of interest. In particular, data showing a trend that characterizes a change due to a system shock are commonly observed over time in biological, medical, health and environmental applications. An example is an experiment on 38 rats  \citep[][also see Section~\ref{s:data}]{reynolds} conducted with an objective to collect information about the state of hypothermia and resuscitation strategy immediately after a 60\% hemorrhage. In practice, hypothermia results in an initial increase in core body temperature, $T_c$, before a decrease takes place. However, critically low $T_c$ may result in a breakdown in the compensatory homeostatic mechanisms \citep{connett, rincon}. Therefore, timely resuscitation from hypothermia is of great clinical interest, which requires (i) the identification of the critical threshold at which $T_c$ starts to decrease, and (ii) the estimation of the decrease rate in $T_c$ after the transition.

\begin{figure}[ht!]\hspace{-0.5cm}
\centering\includegraphics[width=6.2in,height=6.2in]{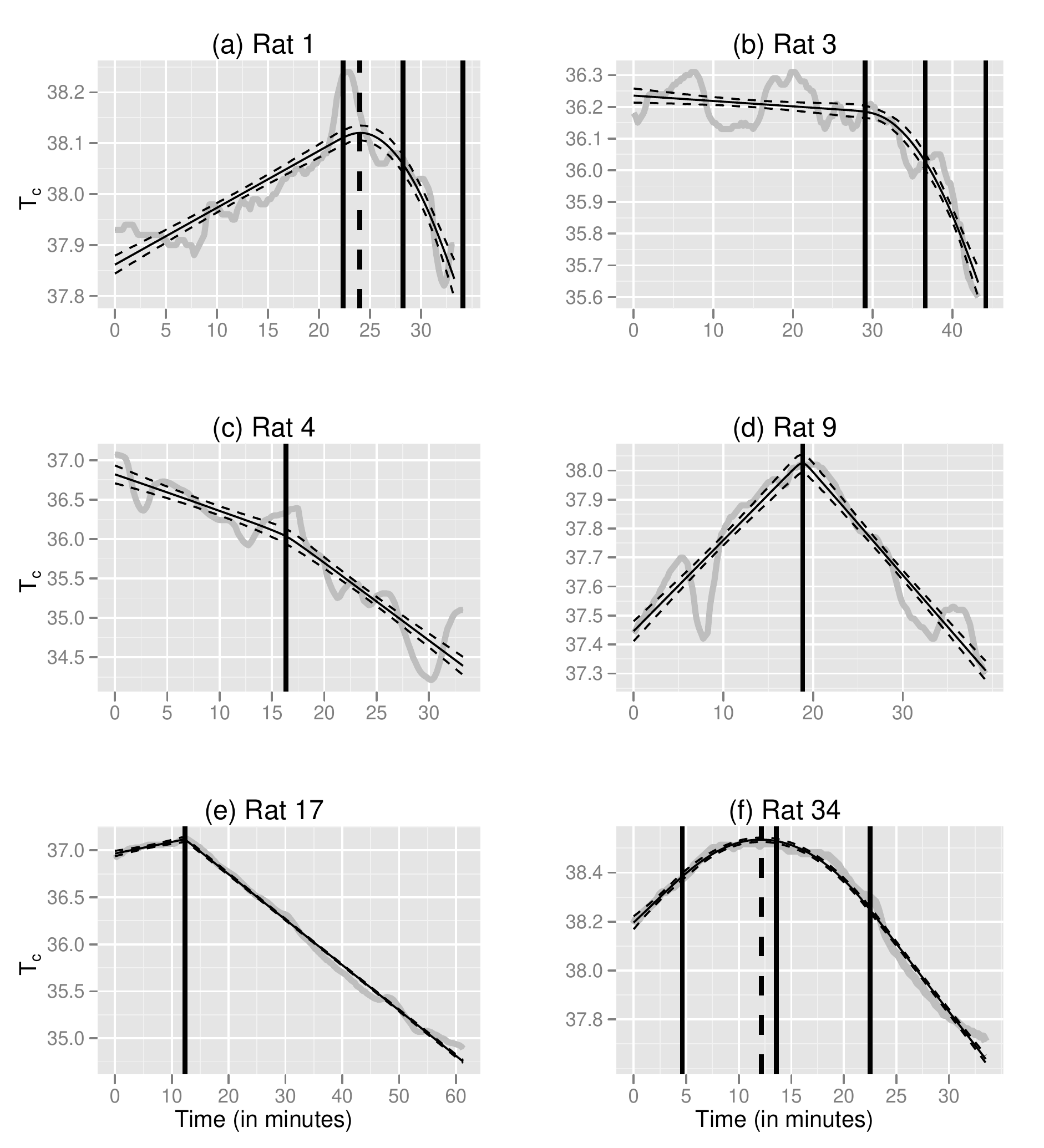}
\caption{Observed profiles (grey curves) and the corresponding individual-specific fitted curves (solid, in black) along with 95\% pointwise credible intervals (dotted curves) for selected rats; fits A and G virtually coincide for each of Rats 4, 9 and 17. Estimated transitions \big(i.e. $\hat{\tau}$ and $\widehat{\tau\pm\gamma}$\big) are marked by solid vertical lines, and estimated CTPs (for Population G) by dotted vertical lines; the CTP estimate is not marked for Rat 3 because the estimated slope of its profile does not change signs. All 38 data profiles appear in supp. material Section S3.}
\label{f:ratindfit}
\end{figure}

Figure~\ref{f:ratindfit} shows six of the 38 temporal profiles of $T_c$ in grey. They are selected to reflect the range of shapes of the 38 profiles. Overall, a similar type of trend is exhibited by all profiles -- roughly linear incoming and outgoing phases are observed at the ends of each profile, with a continuous transition between phases. Some rats exhibit a gradual transition, while others, an abrupt transition. That is, we have samples potentially coming from two populations, labelled, G (gradual) and A (abrupt), respectively, according to the type of transition for the underlying $T_c$ trend. An exception in the figure is Rat 4, which exhibits neither an obviously gradual nor abrupt transition, but rather a seemingly linearly decreasing trend. There are only four of such profiles in the dataset, not adequate for hypothesizing an additional population. Such an investigation could be possible with a sufficiently large dataset.\vspace{0cm}

\begin{figure}[ht!]\vspace{-0.1cm}
\centerline{\includegraphics[width=5.5in,height=2.2in]{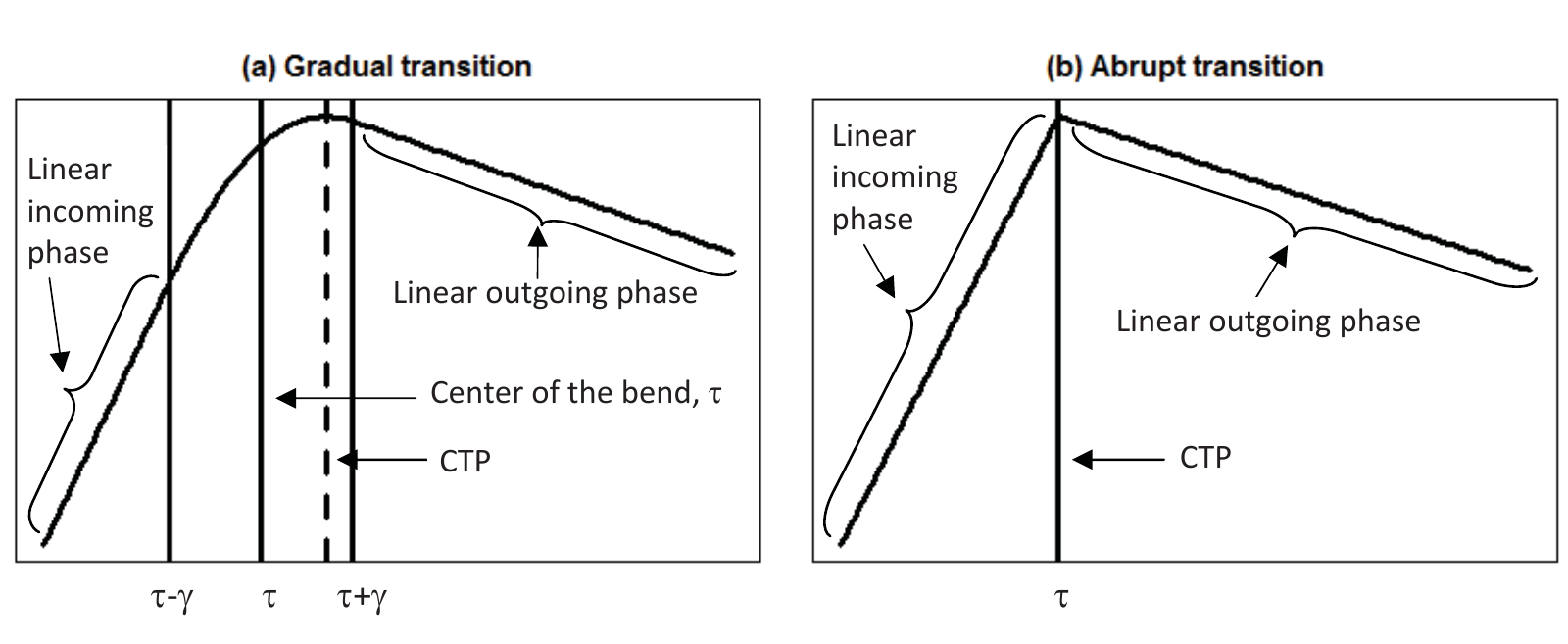}}
\caption{The bent-cable function. (a) A gradual quadratic transition joining two linear segments
(incoming and outgoing). The transition period ranges from $\tau-\gamma$ to $\tau+\gamma$. Any sign change in the slope takes place at the critical time point (CTP). (b) An abrupt transition with $\gamma=0$ yields a broken stick. The change in slope takes place at $\tau$, which is also the CTP.}
\label{f:bentcable}\vspace{0cm}
\end{figure}

Accounting for the possibility of two well-defined populations, A and G, given as few as 38 rats, we develop a statistical framework for modelling these data with particular interest to address (i) and (ii) mentioned above, among related questions concerning therapeutic hypothermia. Our modelling approach is a substantial generalization of a special changepoint model, the \emph{bent cable} \citep{chiu_lockhart_routledge}. It provides flexibility with which inference for the type of transition for each individual is data driven, rather than pre-assumed as a specific type. \citet{chiu_lockhart_routledge} and \citet{chiu_lockhart} developed the bent-cable regression methodology and inference asymptotics to analyze a single data profile showing roughly three phases: incoming and outgoing, both of which are linear, joined by a quadratic bend (Figure~\ref{f:bentcable}(a)). As an extremely sharp bend reduces~the~bent cable to a \emph{broken stick} (Figure~\ref{f:bentcable}(b)), the former encompasses the latter as~a~limiting~case. Although the model is parsimonious and appealing~due~to its simple structure and great interpretability, the authors pointed out that the segmented nature of the model may lead to poor asymptotic approximation in many practical settings involving finite samples.

\citet{khan2009} showed an extension of the bent-cable regression for longitudinal data by explicitly hypothesizing that the sample came from Population G only (henceforth, we will refer to it as Model G), but virtually no methodological details were provided. Their emphasis was on an atmospheric phenomenon that took decades to develop a clear temporal trend. In contrast, in this paper, we investigate two distinct physiological trends that manifest themselves within minutes. For this, we develop our flexible methodology to account for both gradual and abrupt transitions, for which Model G is a simpler~special~case.

The piecewise linear (broken stick) model has been heavily utilized to describe a continuous trend exhibiting at least one abrupt change over time. \citep[e.g.,][] {bell, hall, kiuchi}. However, abruptness of change for all individuals necessarily imposed by the broken stick is unrealistic for the hypothermia experiment, as demonstrated by Figure~\ref{f:ratindfit}. Other quantitative methods that ignore temporal correlation and treat each subject separately are also unsatisfactory. This is because temporal correlation is typically substantial \citep{reynolds_chiu}. Moreover, based on the individual model fits, any inferential statement for the underlying population is at best ad hoc. The use of our flexible bent cable framework presented in this paper advances insight into the important aspect of the quantification of the $T_c$ trend over time by (a) relaxing universal abruptness or gradualness through mixture modelling of piecewise linear and bent cable, (b) allowing proper inference at the population level pooling all individuals in a mixed effects longitudinal framework, which also (c) mitigates practical difficulties with modelling, such as the need for ignoring outlier individuals and for arbitrary truncation of data (as were necessary for \citealt{reynolds_chiu}; see Section~\ref{s:data}). Note the practical significance of (a): if all individuals are assumed to exhibit the same type of transition, then shrinkage towards the population may force an observed profile resembling a broken stick (e.g., Figure~\ref{f:ratindfit}e) to take on a bent-cable fit, and vice versa. Such bias will be demonstrated in Sections~\ref{ss:analysis1} and \ref{s:simulation}. That is, due to shrinkage, the broken stick being a special case of the bent cable (Figure~\ref{f:bentcable}) does not necessarily prevent biased inference for an individual~rat. Thus, property (a) is crucial for proper inference in the context of therapeutic hypothermia.

Although flexible modelling approaches such as penalized spline regression \citep{ruppert} can also handle abrupt and/or gradual changes, the added flexibility in the shape of the fitted model can come at a cost of interpretability. In contrast, we use a model that is simultaneously flexible, interpretable and parsimonious at the population level (see Section~\ref{s:model}), and therefore potentially valuable to many scientific contexts.

In Section~\ref{s:data}, we describe the rat study and outline some substantive research questions about hypothermia therapy. In Section~\ref{s:model}, we present our modelling framework to account for either type of transition through a longitudinal mixture model extension of the single-profile bent-cable regression technique. Additionally, an autoregressive process \citep{box} of order $p$ (AR($p$)), $p\geq 0$, is considered to approximate the within-individual autocorrelation structure. This is all constructed under a Bayesian framework (Section~\ref{s:analysis}), so that the concern about unsatisfactory performance of bent-cable asymptotics is irrelevant. We then apply our method to the aforementioned rat data (Sections~\ref{s:results}) to address the questions regarding hypothermia therapy. In Section~\ref{s:simulation}, simulations (a) demonstrate the importance of hypothesizing both Populations A and G for the rat model, and (b) illustrate that the flexible methodology can perform well with respect to the population regression coefficients even for a misspecified within-individual correlation structure, a fact which is taken into account in analyzing the rat data. We summarize our findings in Section~\ref{s:discuss}.\vspace{0.1cm}

\section{Data and Research Questions}\vspace{0.1cm}
\label{s:data}
Neuronal damage is a common outcome for the survivors of cardiac arrest. Cardiac arrest generally leads to a decrease in the level of oxygen, a condition called anoxia which~our~brain can tolerate for up to 2 to 4 minutes \citep{krause}; irreversible brain damage begins to occur thereafter. In fact, anoxic brain injury is the outcome of a complex process -- ischemia and the subsequent reperfusion together cause enormous biochemical, structural and functional insults that lead to progressive cell destruction, multiorgan dysfunction and neural apoptosis \citep{negovsky}. Hypothermia can protect the brain and heart by attenuating or ameliorating the deleterious temperature-sensitive mechanisms of that process.

Effects of hypothermia on metabolism include a decrease in cerebral blood flow and brain volume, reduction of metabolism, diminution of intracranial pressure, and inhibition of glutamate release and other pathophysiological mechanisms \citep{rincon}. It protects tissue from ischemic damage through this process \citep{gordon}. In contrast, when the body becomes very cold, all physiological systems begin to slow down, eventually to the point that threatens survival. Treatment priorities in such situations include prevention of further cooling and resuscitation \citep{martyn}. Therefore, the main research interest lies in quantification of the transition of $T_c$ to early hypothermia.

Motivated by the above, \citet{reynolds} conducted the aforementioned~rat~experiment (a rat model for humans) to understand hypothermia therapy. Below we summarize the experiment as described by \citet{reynolds_chiu}. Thirty-eight approximately 8-week-old male Long-Evans rats were used in the experiment.~Intraperitoneal~transponders were implanted in the rats to record $T_c$. Followed by a recovery time of 30~--~60 minutes, the rats were hemorrhaged from the carotid catheter with a constraint of a mean arterial pressure threshold of 40 mm Hg. The experiment was continued until the target shed blood volume (60\% of the total blood volume) was achieved, which was then followed by a resuscitation intervention. Core temperature, $T_c$, was logged by automated remote data collection every 15 seconds for the duration of the trial; $T_c$ for each rat was between 127 and 246 time-steps (32 to 62 minutes) long.

\citet{reynolds_chiu} used broken sticks and bent cables to model the rat profiles, treating each as an individual time series. As such, they needed to omit some ``outlying'' profiles that did not obviously follow the shape of either the broken~stick~or the bent cable, and to truncate some other profiles which violated linearity of the incoming or outgoing phase; analyses for only 23 rat profiles were reported. In contrast,~with~our more general mixture methodology (Section~\ref{s:model}) proposed here, we can unify the inference~from~all 38 rats to address questions of broad interest about the underlying rat population, particularly: (a) How long did it take for the $T_c$ trend to show an obvious change because of hypothermia? (b) What were the rates of increase/decrease before and after the change? (c) What was the time point at which the trend went from increasing to decreasing, or vice versa? Like \citet{reynolds_chiu}, we consider data from the start of hemorrhage until resuscitation intervention. On the other hand, an obvious advantage of our method is that properly accounting for the longitudinal context allows pooling of information from the entire sample, overcoming computational difficulties due to the apparent violation of the shape of the broken stick or bent cable for certain individuals.

\section{The Flexible Mixture Longitudinal Bent-Cable Model}
\label{s:model}
When $m$ individuals can be regarded as having been randomly selected from some population and repeated measurements are observed for each individual, it is useful to unify information from all $m$ to aid the understanding of the population as well as subject-specific behaviour. For the $i^{th}$ individual $(i=1,2,\ldots,m)$, let there be $n_i$ measurements, and let $t_{ij}$ denote the $j^{th}$ measurement occasion, $j=1, 2, \dots, n_i$. We model the corresponding response at time $t_{ij}$, denoted by $y_{ij}$, by the relationship
\begin{equation}
y_{ij}=f(t_{ij},\boldsymbol{\theta}_i)+\epsilon_{ij}
\label{e:reg}
\end{equation}
where $\boldsymbol{\theta}_i$ is the vector of regression coefficients for the $i$th individual, $f(\cdot)$ is a function of $t_{ij}$ and
$\boldsymbol{\theta}_i$ to characterize the trend of the subject-specific data, and $\epsilon_{ij}$ represents the random error component, which accounts for measurement error and possibly additional within-individual error.

For the types of data under consideration, in light of the apparent three phases -- linear incoming and outgoing, and the adjoining curved transition -- we characterize the individual profiles by the bent-cable function \citep{chiu_lockhart_routledge}, given by
\begin{equation}
f(t_{ij},\boldsymbol{\theta}_i)=\beta_{0i}+\beta_{1i}t_{ij}+\beta_{2i} q(t_{ij},\boldsymbol{\alpha}_i),
\label{e:f}\vspace{0.45cm}
\end{equation}
where\vspace{-1.3cm}
\begin{equation}
q(t_{ij},\boldsymbol{\alpha}_i)=\frac{(t_{ij}-\tau_i+\gamma_i)^2}{4\gamma_i} \mathbf{1}\{|t_{ij}-\tau_i|\leq\gamma_i\}+ (t_{ij}-\tau_i)\mathbf{1}\{t_{ij}-\tau_i > \gamma_i\}
\label{e:q}
\end{equation}
with $\boldsymbol{\beta}_i=(\beta_{0i},\beta_{1i},\beta_{2i})'$ and $\boldsymbol{\alpha}_i=(\gamma_{i},\tau_{i})'$ being the vectors of linear and transition coefficients, respectively, and $\boldsymbol{\theta}_i=(\boldsymbol{\beta}_i',\boldsymbol{\alpha}_i')'$. For each individual $i$, $\beta_{0i}$ and $\beta_{1i}$ are, respectively, the intercept and slope of the incoming phase; $\beta_{1i}+\beta_{2i}$, the slope of the outgoing phase; and $\tau_{i}$ and $\gamma_{i}$, the transition parameters which  represent the center and half-width of the bend, respectively. Henceforth, we will denote $f(t_{ij},\boldsymbol{\theta}_i)$ and $q(t_{ij},\boldsymbol{\alpha}_i)$ by $f_{ij}$ and $q_{ij}$. Note that $\gamma_i=0$ reduces the bent cable to a broken-stick model for which $q_{ij}=(t_{ij}-\tau_i)\mathbf{1}\{t_{ij}-\tau_i > 0\}$ (Figure~\ref{f:bentcable}(b)).

The critical time point (CTP), as defined by \citet{chiu_lockhart}, is the time at which the slope of the bent cable changes signs (Figure~\ref{f:bentcable}). Thus, for a gradual transition, the CTP is $\tau_i-\gamma_i-2\beta_{1i}\gamma_i/\beta_{2i}$. Note that this formula is not meaningful when the slope of the cable does not change signs. When $\gamma_i=0$, any sign change of the slope occurs at the point $\tau_i$, the CTP for an abrupt transition.

We consider a hierarchical mixed-effects modelling framework and regard $\boldsymbol{\theta}_i$ as random, through which we can obtain useful information regarding the questions: (1) How does the response change over time (a) individually and (b) at the population level? and (2) Do different individuals experience different patterns of change? Question (1)(a) characterizes each individual's pattern of change over time (commonly called \emph{within-individual} or Level 1 variation), and (2) addresses the association between patterns of change (commonly called \emph{between-individual} or Level 2 variation). Additionally, there is a third level for Bayesian inference, which quantifies prior knowledge for (1) and (2).

The framework as described thus far constitutes the longitudinal bent-cable model of \citet{khan2009}. For a mixed-effects model, it is parsimonious in the sense that the underlying population model is the bent cable which has only five fixed-effects regression coefficients. However, undesirable shrinkage issues as described in Section~\ref{s:introduction} are evident when their framework is directly applied to the rat data (see Section~\ref{ss:analysis1}). Thus, to avoid estimation bias due to shrinkage, we further assume that\vspace{-0.1cm}
\begin{list}{\labelitemi}{\leftmargin=2.25em \itemindent=-0em}
\item[\textbf{A1.}] each individual $i$ potentially comes from one of two populations: Population A for which $\gamma_i=0$ and Population G for which $\gamma_i>0$; and\vspace{-0.2cm}
\item[\textbf{A2.}] each individual has probability $\omega$ to have come from Population G (and, hence, probability $1-\omega$ from Population A).\vspace{-0.2cm}
\end{list}
Conditional on the random effects $\boldsymbol{\theta}_i$'s, the sets of repeated measurements $\{y_{i1},y_{i2},\ldots,$ $y_{in_i}\}$ and $\{y_{k1},y_{k2},\dots,y_{kn_k}\}$ are assumed independent for $i \neq k$. To account for additional serial correlation among $y_{ij}$'s remaining after what has been accounted for by the $\boldsymbol{\theta}_i$'s, we assume at Level 1 that $\epsilon_{ij}$'s follow a stationary AR($p$) process with a common~$p$:\vspace{0cm}
\begin{equation}
\epsilon_{ij}=\phi_1 \epsilon_{i,j-1}+\phi_2 \epsilon_{i,j-2}+\ldots+\phi_p \epsilon_{i,j-p}+v_{ij},
\label{e:ap}\vspace{0cm}
\end{equation}
where $\boldsymbol{\phi}=(\phi_1,\phi_2,\ldots,\phi_p)'$ is the vector of AR($p$) parameters, and $[v_{ij}|\sigma_i^2] \sim N(0,\sigma_i^2)$ for all $j=1,2,\ldots,n_i$. Furthermore, we consider a conditional likelihood framework for Level 1, where the initial $p$ observations for each $i$, $\mathbf{y}_i^{(1)}=(y_{i1},y_{i2},\ldots,y_{ip})'$, are~treated~as known, whereas $\mathbf{y}_i^{(2)}=(y_{i,p+1},y_{i,p+2},\ldots,y_{i,n_i})'$ are random. This framework for $\mathbf{y}_i^{(1)}$~and $\mathbf{y}_i^{(2)}$ was also considered by \citet{chiu_lockhart} for frequentist bent-cable regression for a single profile, and by \citet{chib} in a Bayesian approach for linear regression.

Assumptions \textbf{A1} and \textbf{A2}, together with Equations (\ref{e:reg})-(\ref{e:ap}), constitute~our \emph{flexible mixture longitudinal bent-cable} model. Letting:
\begin{equation*}
x_{ij}=t_{ij}-\sum_{k=1}^p{\phi_k}~t_{i,j-k},~~ r_{ij}=q_{ij}-\sum_{k=1}^p{\phi_k~q_{i,j-k}},\vspace{0.4cm}
\end{equation*}
and\vspace{-0.9cm}
\begin{equation*}
\mu_{ij}=\beta_{0i} (1-\sum_{k=1}^p{\phi_k})+\beta_{1i}x_{ij}+\beta_{2i}r_{ij}+\sum_{k=1}^p{\phi_k~y_{i,j-k}}\vspace{0.2cm}
\end{equation*}
for $j=p+1,p+2, \ldots, n_i$, our choices of distributions for the relevant quantities allow us to rewrite the model as \vspace{-0.2cm}
\begin{equation}
[y_{i,p+t}|y_{i,t}, y_{i,t+1}, \ldots, y_{i,p-1+t}, \boldsymbol{\theta}_i,\boldsymbol{\phi},\sigma_{i}^2] \sim N(\mu_{i,p+t},\sigma_{i}^2)\ \forall\ t=1,\ldots,n_i-p,
\label{e:level1}\vspace{0cm}
\end{equation}
\begin{equation}
\left.
\begin{array}{c}\doublespacing
[\boldsymbol{\beta}_i|\boldsymbol{\mu}_\beta,\Sigma_\beta] \sim N_3(\boldsymbol{\mu}_\beta,\Sigma_\beta),\vspace{0.2cm}\\
g(\boldsymbol{\alpha}_i|I_i)=(1-I_i)~LN(\tau_i|\mu_{\tau_A},\sigma_{\tau_A}^2)+
I_i~LN_2(\boldsymbol{\alpha}_i|\boldsymbol{\mu}_\alpha,\Sigma_\alpha),\vspace{0.2cm}\\
I_i \sim BER(\omega)
\end{array}\right\},
\label{e:level2}
\end{equation}
\begin{equation}
\left.
\begin{array}{c}\doublespacing
[\boldsymbol{\mu}_\beta|\mathbf{h}_1,\mathbb{H}_1] \sim N_3(\mathbf{h}_1,\mathbb{H}_1),~
\left[\boldsymbol{\mu}_\alpha|\mathbf{h}_2,\mathbb{H}_2\right] \sim N_2(\mathbf{h}_2,\mathbb{H}_2),\vspace{0.15cm}\\
\left[\boldsymbol{\phi}|\mathbf{h}_3,\mathbb{H}_3\right] \sim N_p(\mathbf{h}_3,\mathbb{H}_3),~
\left[\mu_{\tau_A}|a_0,a_1\right] \sim N(a_0,a_1),\vspace{0.15cm}\\
\left[\Sigma_\beta^{-1}|\nu_1,\mathbb{A}_1\right] \sim W\big(\nu_1,(\nu_1 \mathbb{A}_1)^{-1}\big),~
\left[\Sigma_\alpha^{-1}|\nu_2,\mathbb{A}_2\right] \sim W\big(\nu_2,(\nu_2\mathbb{A}_2)^{-1}\big),\vspace{0.15cm}\\
\left[\sigma_{\tau_A}^{-2}|b_0, b_1\right] \sim G(\frac{b_0}{2},\frac{b_1}{2}),~
\left[\sigma_{i}^{-2}|d_0,d_1\right] \sim G(\frac{d_0}{2},\frac{d_1}{2}),\vspace{0.15cm}\\
\left[\omega|c_0,c_1\right] \sim B(c_0,c_1)
\end{array}\right\},
\label{e:level3}\vspace{0.4cm}
\end{equation}
where $\boldsymbol{\mu}_\beta \equiv (\mu_0,\mu_1,\mu_2)'$ and $\Sigma_\beta$ are, respectively, the mean and covariance of $\boldsymbol{\beta}_i$; $\mu_{\tau_A}$ and $\sigma_{\tau_A}^2$ are the mean and variance of $\log{(\tau_i)}$ for Population A, $\boldsymbol{\mu}_\alpha\equiv (\mu_\gamma,~\mu_\tau)'$ and $\Sigma_\alpha$ are the mean and covariance of $\log{(\boldsymbol{\alpha}_i)}$ for Population G; and $N_p$,~$LN_p$,~$BER$,~$W$,~$G$~and~$B$ stand for $p$-variate normal, $p$-variate lognormal, Bernoulli, Wishart, gamma and beta distributions,~respectively. Levels~1~and~2 are (\ref{e:level1}) and (\ref{e:level2}), and Level 3 is (\ref{e:level3})~with~the~hyperparameters assumed known (see supp. material Section S1). Note that the distribution of $\boldsymbol{\alpha}_i$ is a mixture of a univariate and a bivariate lognormal distribution corresponding to assumption \textbf{A1}; $\boldsymbol{\alpha}_i=(\gamma_i,\tau_i)'$ for $I_i=1$, and $\boldsymbol{\alpha}_i=[\tau_i]$ for $I_i=0$ due to a deterministic $\gamma_i=0$.

\section{Rat Data Analysis: Bayesian Inference and Implementation}\vspace{0.1cm}
\label{s:analysis}
\subsection{The Longitudinal Bent-Cable Model}
\label{ss:analysis1}
The assumption that the samples come from Population A only is, perceivably, a restrictive and unrealistic assumption for a physiological phenomenon. As the existing framework by \citet{khan2009} (Model G) allows an arbitrarily small $\gamma_i>0$ for each $i$, we first applied it to our rat data to generalize this restrictive assumption. We observed an unusually large upper limit for the 95\% credible interval for $(\Sigma_\alpha)_{11}$, i.e.,  the variance of $\gamma_i$. This impracticality can be explained by noting that the presence of any rat $i$ whose posterior draws for $\gamma_i$ are arbitrarily small (e.g., $<10^{-3}$) can substantially inflate the corresponding posterior draws for $(\Sigma_\alpha)_{11}$. As several rats exhibit a virtually abrupt transition while others do not, this resulted in an unreasonable estimate of $(\Sigma_\alpha)_{11}$.

\subsection{The Flexible Mixture Longitudinal Bent-Cable Model}
\label{ss:analysis2}
Thus,~there is practical~need to~generalize Model~G~by further~extending it to a mixture~of~A and G as described in Section~\ref{s:model}. As we explain below, the analysis using our mixture model provides strong evidence that supports the existence of not just A nor just G for the rat study.

Bayesian inference is carried out by \emph{Markov chain Monte Carlo} (MCMC), where~we sample from the posterior distribution by the Metropolis within Gibbs algorithm \citep{smith}. For implementation, we work out the full conditional for each parameter (see supp. material Section S2). We employ the Metropolis algorithm to draw samples of $\boldsymbol{\alpha}_i$, the sole parameter for which the full conditional can be expressed only up to a proportionality constant. The full conditional for $\boldsymbol{\phi}$ is Gaussian; we take the~proportion~of draws (from the full conditional for $\boldsymbol{\phi}$) for which stationarity is satisfied as an~estimate~of the conditional probability of stationarity for the AR process \citep{chib}.~We~consider~several models assuming $\{\epsilon_{ij}\}$ to be AR($p$) for $p=0,1, \ldots$, and choose the one for~which~the estimate of the deviance information criterion (DIC) is minimum \citep{spieg}.

Since our assumption for the~$\boldsymbol{\alpha}_i$'s involves lognormal distributions, we can use Level~2 medians,~namely~$\mbox{\fontsize{10}{12}\selectfont $\mathcal{M}_\gamma$}$$\mbox{\fontsize{11}{13}\selectfont $\equiv \exp{\{\mu_{\gamma}\}}$}$~and~$\mbox{\fontsize{10}{12}\selectfont $\mathcal{M}_\tau$}$$\mbox{\fontsize{11}{13}\selectfont $\equiv \exp{\{\mu_{\tau}\}}$}$ for Population G and $\mbox{\fontsize{10}{12}\selectfont $\mathcal{M}_{\tau_A}$}$$\mbox{\fontsize{11}{13}\selectfont $\equiv \exp{\{\mu_{\tau_A}\}}$}$ for Population A, to describe the transition locations. We can also use Level 2 standard deviations of $\gamma_i$ and $\tau_i$ for G, namely $\mbox{\fontsize{10}{12}\selectfont$\mathcal{S}_\gamma$}$$\mbox{\fontsize{11}{13}\selectfont $\equiv \sqrt{\exp{\{2\mu_\gamma+(\Sigma_\alpha)_{11}\}}\times [\exp{\{(\Sigma_\alpha)_{11}\}}-1]}$}$ and $\mbox{\fontsize{10}{12}\selectfont$\mathcal{S}_\tau$}$$\mbox{\fontsize{11}{13}\selectfont $\equiv \sqrt{\exp{\{2\mu_\tau+(\Sigma_\alpha)_{22}\}}\times [\exp{\{(\Sigma_\alpha)_{22}\}}-1]}$}$ to describe the between-individual variability of these transition parameters. Posterior means or medians of $\mathcal{M}$s and $\mathcal{S}$s can be easily approximated using the MCMC samples.

We proceed to analyze the aforementioned rat data using our flexible~mixture bent-cable approach. We denote time by~$t_{ij}$,~$j=1,2,\ldots,n_i$, where $t_{i1}=0$~refers to the starting point of the study for rat $i$ $(i=1,2,\ldots,38)$, and each subsequent time increment is~15~seconds. Any parameter estimate~(Level~1~or~2)~is based~on the posterior mean or median, depending on the extent of asymmetry of the corresponding marginal posterior density. Note that $\boldsymbol{\theta}_i$ has its own posterior distribution, inducing a posterior distribution for the bent cable~$f_{ij}$~at each observed~$t_{ij}$. Therefore, we regard the MCMC sample mean of $f_{ij}$ as the fitted value $\hat{f}_{ij}$. Individual-specific fitted curves are then interpolated based on the $\hat{f}_{ij}$ values; see Appendix~\ref{s:fitted_values}. A fitted population curve is produced based on the estimates of the theoretical medians for $\boldsymbol{\beta}_i$ and $\boldsymbol{\alpha}_i$ from Level 2. Similarly, we define the CTP for Population G as $\mbox{\fontsize{11}{13}\selectfont $\mathcal{M}_{\tau}-\mathcal{M}_{\gamma}-2\mu_{1}\mathcal{M}_{\gamma}/\mu_{2}$}$; thus, we use the posterior mean of this expression to make inference for this CTP\hspace{0.3mm}. Estimates for the other parameters for Level 2 Population A/G theoretical medians/standard deviations are produced similarly.

\section{Results}\vspace{-0.1cm}
\label{s:results}
Initially, we consider several flexible bent-cable models based on the degree of within-individual dependency among the repeated measurements, which is assumed through an AR($p$) process for $p=0, 1,2, 3$. Model selection procedure reveals a smallest~DIC~for~the AR(0) assumption. Fixing $p=0$, we then analyzed the data using Models G and A (i.e., assuming that the sample arises from Population G only and Population A only), and observed the smallest DIC for the proposed flexible model; see Appendix~\ref{s:model_selection} for details. Therefore, we report here the results for the flexible model with~AR(0)~within-individual~noise.

\begin{figure}[ht!]
\centerline{\includegraphics[width=4.5in,height=3.4in]{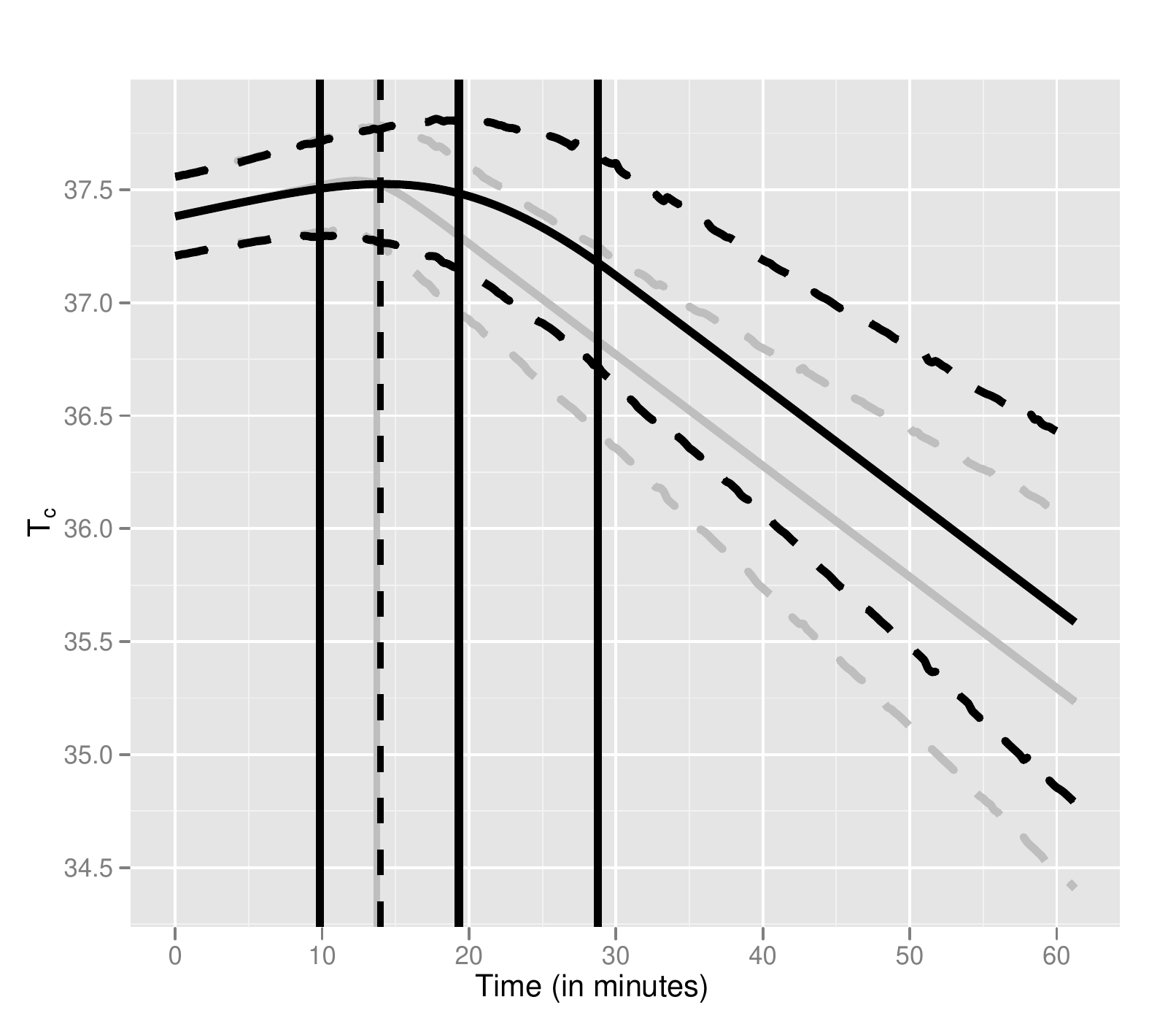}}
\caption{Fitted population curves (solid) with 95\% pointwise credible intervals (dotted curves) for the two populations (grey for A and black for G). The model fit is produced assuming conditional within-individual independence. The estimated transition for G \big(i.e. $\widehat{\mbox{\fontsize{7}{9}\selectfont $\mathcal{M}_\tau\pm\mathcal{M}_\gamma$}}$\big) is marked by solid black vertical lines, and that for A \big(i.e. $\mbox{\fontsize{7}{9}\selectfont $\widehat{\mathcal{M}}_{\tau_A}$}$\big) by the grey vertical line. The estimated CTP for G is indicated by the dotted vertical line. (See supp. material Figure~\ref{f:ratpopfit1} in Section S3 for the same figure but overlaid with all 38 profiles.)}
\label{f:ratpopfit}\vspace{-0.3cm}
\end{figure}

Some posterior characteristics of parameters for the two populations are given in Table~\ref{t:sumfitrat}, and the population fitted curves are displayed in Figure~\ref{f:ratpopfit}.~The~posterior~mean for~$\omega$~is~$0.39$, suggesting that about~39\%~of the rats~belong to Population G which exhibits a gradual change in $T_c$. Posterior means for $\mbox{\fontsize{9}{11}\selectfont$\mathcal{M}_{\tau}\pm\mathcal{M}_{\gamma}$}$ are $10.11$ and $29.03$~minutes,~implying that the population~transition begins~approximately $10.11$ minutes from the time of hemorrhage~and lasts~for about $18.92$ minutes, followed by a significant linear decrease at the rate of $0.013^oC$ per 15 seconds (the posterior mean for $\mu_1+\mu_2$ is $-0.013$ with 95\% credible interval $(-0.016,-0.008)$). The remaining 61\% of the rats, approximately, exhibit an abrupt linear decrease at the same rate from the transition time point. We also see a significant linear increase in population $T_c$~at the rate of $0.003^oC$ per 15 seconds in the incoming phase~\big(95\% credible interval of the incoming slope is $(0.001, 0.006)$\big). Moreover, virtually identical metabolic~thresholds associated with a breakdown in the compensatory mechanisms for the two populations are observed (see Figure~\ref{f:ratpopfit}): posterior means for Population G and A CTPs are $14.28$ and $13.89$ minutes, respectively (Table~\ref{t:sumfitrat}). Thus, for G, the drop in $T_c$ started at approximately $14.28$ minutes after hemorrhage, and $13.89$ minutes~for~A.

\begin{table}[ht!]
\caption{\small Posterior summaries for the two populations of rats assuming AR(0) noise: posterior means for the population slope parameters ($\mu_1$ and $\mu_2$) are in ``per 15 seconds'' and those for the population transitions are in minutes.} \centering \small
\begin{threeparttable}
\begin{tabular*} {\textwidth}{@{\extracolsep{\fill}} l c c c}
\hline \smallskip
  &&Posterior& 95\% credible\\
  &&mean& interval\\
  \hline \vspace{0.15cm}
  $\omega$~~~ (Probability of being from G) &&$0.39$ & $(0.23, 0.55)$\\ \vspace{0.15cm}
  $\mu_0$~~~ (Incoming intercept) &&$37.38$ & $(37.21, 37.56)$\\ \vspace{0.15cm}
  $\mu_1$~~~ (Incoming slope) && $0.003$& $(0.001, 0.006)$\\  \vspace{0.15cm}
  $\mu_2$~~~ (Difference between incoming and outgoing slopes) && $-0.016$& $(-0.020, -0.011)$\\ \vspace{0.15cm}
  $\mathcal{M}_{\tau_A}$~~ (Population CTP for A)&&$13.89$ & $(10.59, 17.34)$ \\ \vspace{0.15cm}
  $\mathcal{M}_{\gamma}$~~~ (Half-width of the bend for G)&&$9.46$ & $(5.45, 13.62)$ \\ \vspace{0.15cm}
  $\mathcal{M}_{\tau}$~~~ (Center of the bend for G)&&$19.57$ & $(13.67, 25.30)$ \\ \vspace{0.15cm}
  $\mathcal{M}_{\tau}\pm\mathcal{M}_{\gamma}$~~~ (Transition period for G) &&10.11 to 29.03 &$-$\\ \vspace{0.15cm}
  $\mathcal{M}_{\tau}-\mathcal{M}_{\gamma}-2\mu_{1}\mathcal{M}_{\gamma}/\mu_{2}$~~~ (Population CTP for G) && $14.28$& (6.33, 21.84)\\
  \hline
\end{tabular*}
\end{threeparttable}
\label{t:sumfitrat}\vspace{0cm}
\end{table}

Figure~\ref{f:ratindfit} shows examples of the individual fitted curves.~For all but~one~of~the six rats displayed (and all but a total of four out of the entire sample of 38 rats), the fits appear very reasonable as the observed data closely agree with the respective fitted lines, and the estimated transitions \big($\hat{\tau}$ and $\widehat{\tau\pm\gamma}$\big) demonstrate that our methodology picks up the two~types~of transition adequately.  The remaining rat (Figure~\ref{f:ratindfit}(c)) appears to be unusual, exhibiting linearly decreasing trends throughout (recall Section \ref{s:introduction}); again, this is one of four rats among the 38 who do not cleanly fall into either population.  With our methodology, these four are estimated to have arisen from Population A. Given our small dataset, we do not consider a potential third population to avoid overfitting.

The posterior characteristics of the theoretical standard deviations and correlations are given in Table~\ref{t:sumvariancerat}. Since the biological conditions of different rats should vary to some extent, we can expect some variation in $T_c$'s at the time of administering hemorrhage. This is reflected in the estimate of the standard deviation of $\beta_{0i}$, which is $0.535$. After administering hemorrhage, we see very little variation in the slope parameters (estimated standard deviations for $\beta_{1i}$ and $\beta_{1i}+\beta_{2i}$ are $0.008$ and $0.011$, respectively), that is, all rats exhibit very similar rates of increase/decrease before/after the transition period. Significant negative correlation between $\beta_{1i}$ and $\beta_{2i}$ \big($\widehat{corr}(\beta_{1i},\beta_{2i})=-0.476$ with 95\% credible interval $(-0.711,-0.204)$ which excludes zero\big) indicates that the steeper the incoming slope going up, the bigger the drop in slope for the outgoing phase.

\begin{table}[ht!]
\caption{\small Rat data analysis -- posterior summaries of the standard deviations and correlations associated with $\Sigma_\beta$, $\Sigma_\alpha$ and $\sigma_{\tau_A}^2$; posterior medians for the standard deviations of the linear parameters ($\beta_{1i}$ and $\beta_{2i}$) are in ``per 15 seconds'' and those for the transition parameters ($\gamma_i$ and $\tau_i$) are in minutes.} \centering \small
\begin{threeparttable}
\begin{tabular*}{\textwidth}{@{\extracolsep{\fill}} l c c c}
\hline \smallskip
  &&Posterior & 95\% credible\\
  &&median& interval\\
  \hline \vspace{0.15cm}
  S.D. of $\beta_{0i}$ &&$0.535$ & $(0.423, 0.669)$\\ \vspace{0.15cm}
  S.D. of $\beta_{1i}$ &&$0.008$ & $(0.006, 0.010)$\\ \vspace{0.15cm}
  S.D. of $\beta_{2i}$ &&$0.123$ & $(0.010, 0.016)$\\ \vspace{0.15cm}
  S.D. of $\beta_{1i}+\beta_{2i}$ &&$0.011$ & $(0.009, 0.014)$\\ \vspace{0.15cm}
  Corr. between $\beta_{0i}$ and $\beta_{1i}$  &&$0.023$ & $(-0.296, 0.343)$\\  \vspace{0.15cm}
  Corr. between $\beta_{0i}$ and $\beta_{2i}$ &&$-0.001$ & $(-0.323, 0.319)$\\ \vspace{0.15cm}
  Corr. between $\beta_{1i}$ and $\beta_{2i}$ &&$-0.476$ & $(-0.711, -0.204)$\\ \vspace{0.15cm}
  S.D. of $\gamma_{i}$ for Population G &&$7.643$ & $(2.967,19.068)$\\  \vspace{0.15cm}
  S.D. of $\tau_{i}$ for Population G &&$10.172$ & $(4.728,20.928)$\\  \vspace{0.15cm}
  Corr. between $\gamma_i$ and $\tau_{i}-\gamma_i$ for Population G &&$-0.815$ & $(-0.976,-0.602)$\\  \vspace{0.15cm}
  S.D. of $\tau_{i}$ for Population A&&$9.312$ & $(5.030,16.200)$\\
  \hline
\end{tabular*}
\end{threeparttable}
\label{t:sumvariancerat}
\end{table}

From Table~\ref{t:sumvariancerat}, we see considerable variability in the times to maximal $T_c$ and the variability in the times to transition zones. This fact is reflected in the posterior medians for the standard deviations of $\gamma_i$ and $\tau_i$ for Population G ($7.643$ and $10.172$ minutes, respectively), and of $\tau_i$ for Population A ($9.312$ minutes). We also see $\widehat{corr}(\gamma_{i},\tau_{i}-\gamma_i)=-0.816$ with 95\% credible interval $(-0.976, -0.602)$, which excludes zero. Significant negative correlation between $\gamma_i$ and $\tau_i-\gamma_i$ indicates that for individuals from Population G, the sooner the gradual transition takes place, the wider the transition zone so that there will be a delayed linear drop in the outgoing phase, and vice versa.

In~summary, our analysis~yields the following~points of clinical~interest: (a)~about~61\%~of the rats exhibit an abrupt linear drop in $T_c$ during hemorrhage, whereas the remaining 39\%, approximately, exhibit a gradual transition followed by a linear drop; (b) all rats are from populations that show approximately the same metabolic threshold (about $14$ minutes after hemorrhage) associated with a breakdown in the compensatory mechanisms; (c) either population shows a significant increase of $T_c$ followed by a significant decrease; (d) all the rats exhibit very similar rates of increase and decrease in $T_c$ before and after the transition period, respectively; (e) there is a considerable amount of between-rat variability in the times to maximal $T_c$ and transition zones; (f) the sooner the gradual transition takes place, the wider the transition zone, and vice versa; (g) although assuming within-subject conditional independence may be unrealistic for some problems, we demonstrate in Scenarios 2 and 3 in the next section that points (a)-(f) above should be reasonably robust to this assumption.

\section{Simulations}\vspace{-0.15cm}
\label{s:simulation}
First, we supplement the motivation for our mixture methodology as seen in Section~\ref{ss:analysis1}. That is, we show the importance of hypothesizing Population A in addition to G in a more general context, despite that the abrupt broken stick is the limiting case of the smooth bent cable. To this end, we present Scenario 1, where we fit Model G when, in reality, both Populations A and G exist, with G heavily dominating A: (a) $\omega=0.90$ and (b) $\omega=0.95$. In both 1a and 1b, \{$\epsilon_{ij}$\} is an AR(1) process with $\phi=0.70$, where $p=1$ is treated as known when fitting Model G.

Second, to illustrate the fact that our flexible methodology can perform well with respect to the population regression coefficients even for a misspecified correlation structure for \{$\epsilon_{ij}$\}, we present Scenarios 2 and 3, where \{$\epsilon_{ij}$\} has AR(1) or AR(2) structure. In each case, we analyze the data assuming $p=0,1,$ and $2$, and that the samples come from two potential populations (A and G). In all the scenarios, we take $m=20$, $n\equiv n_i=150$ for $i=1,2,\ldots,m$ and $t_{ij}=j-1$ for $j=1,2,\ldots,n$. Model parameter values were chosen to allow reasonable generalization, and are given in Tables~\ref{t:pop1}~-~\ref{t:innovation2} in supp. material Section~S\ref{s:simresults}.

For each simulation, $500$ data sets are generated, and $100,000$ MCMC iterations are used to approximate posterior distributions per set. Posterior summaries are averaged over the 500 sets for each parameter, and the coverage probability of 95\% credible intervals (proportion of such credible intervals out of 500 that capture the truth) is calculated.
\subsection{Results for Scenario 1}\vspace{-0.1cm}
\label{ss:scenario1}
Numerical results are tabulated in Tables~\ref{t:pop1}~-~\ref{t:innovation1} in supp. material Section S4.1. We see that Model G performs well with respect to all but one population regression coefficient, $\mu_\gamma$, the bend's half-width (for Population G). Specifically, the average of the posterior means for each parameter except $\mu_\gamma$ is close to the true parameter value, and the corresponding coverage probabilities are all reasonably close to the nominal $0.95$. When $\omega=0.90$, we see underestimation and under coverage for $\mu_\gamma$. This can be explained by noting that the average of the posterior means for each $\gamma_i$ is expected to be approximately zero for profiles that originate from A; this leads to underestimation of the population counterpart $\mu_\gamma$. Note that if we would model this data set using our flexible methodology, $\mu_\gamma$ would represent only the profiles that originate from G, and hence, underestimation for $\mu_\gamma$ would not be expected, and coverage for $\mu_\gamma$ would be close to the nominal $0.95$. Indeed, this is evident from the results for $\omega=0.95$ (fewer abrupt profiles than $\omega=0.90$):~the average of the posterior means for $\mu_\gamma$ is $50\%$ closer to the true value of $3$, and also~the coverage is $67\%$ closer to the nominal $0.95$ (see supp. material Section S4.1).

Details about $\Sigma_\beta$, $\Sigma_\alpha$, and $\sigma_i^2$'s also appear in Section S4.1. The main conclusion is that (\rmnum{1}) the estimation of $(\Sigma_\alpha)_{11}$ and $(\Sigma_\alpha)_{22}$ is more accurate for $\omega \approx 1$, and (\rmnum{2}) misspecifying the model as Model G, when in reality both populations A and G exist, negligibly affect the estimates of $\sigma_i^2$'s or $\Sigma_\beta$.

The above simulation results demonstrate the importance of modelling Population A distinctly from G using our flexible methodology to analyze data that resemble those from the rat experiment; note the rat experiment required an even more extreme need for a mixture, with 95\% credible interval for $\omega$ being $(0.23, 0.55)$.

\subsection{Results for Scenarios 2 and 3}
\label{ss:scenario23}
Numerical results for Scenarios 2 and 3 are given in Tables~\ref{t:pop2}~-~\ref{t:innovation2} in supp. material Section S4.2. Our methodology performs well for both scenarios with respect to the population characteristics: averages of posterior means are all close to the true parameter values, and coverage probabilities (from $0.92$ to $0.99$) are all reasonably close to the nominal $0.95$. This suggests that our Bayesian inference for population characteristics is robust to ignoring certain types of serial correlation.

Details about $\Sigma_\beta$, $\Sigma_\alpha$, $\sigma^2_{\tau_A}$, and $\sigma_i^2$'s also appear Section S4.2. The~main conclusion is that (\rmnum{1}) the estimation of $\Sigma_\beta$, $\Sigma_\alpha$, and $\sigma^2_{\tau_A}$ is quite accurate for correctly specified models, though underspecifying $p$ may result in under coverage for $(\Sigma_\alpha)_{11}$, $(\Sigma_\alpha)_{22}$ and $\sigma^2_{\tau_A}$, and (\rmnum{2}) an underspecified $p$ leads to overestimation of $\sigma_{i}^2$. In particular, we observe very poor coverage for $\sigma_i^2$ if we incorrectly analyze a data set by an AR(0) assumption when, in reality, it exhibits serial correlation over time. Although poor coverage may~not~be ideal in certain cases, of primary practical concern in the rat analysis is the inference for the population regression coefficients, for which our methodology demonstrates robustness.

\section{Conclusion}\vspace{-0.1cm}
\label{s:discuss}
Induced hypothermia potentially saves lives under physiological trauma. Yet, without extreme care, it can also threaten survival. Therefore, controlled administration of hypothermia is of paramount importance. In this article, we developed the flexible mixture bent-cable framework to quantify the transition of core body temperature, $T_c$, during induced hypothermia in a rat model for humans. Our analysis reveals important clinical information that can be very valuable in administering hypothermia therapy. Aside from crucial information at the population level, another aspect of our longitudinal framework which clinicians would find valuable is the inference for the temporal trend exhibited by individual observational units: the current inference for an individual may provide guidelines to future administration of hypothermia therapy to the same individual.

The most appealing feature of our method may be its greatly interpretable parameters, and that useful information can be obtained at the small cost of estimating very few fixed-effects regression coefficients. Moreover, pooling information from many individuals leads to shrinkage, so that mild deviations of observed profiles from the broken-stick/bent-cable structure do not hinder model fitting; in contrast, deviations considered mild can render the single-profile bent-cable regression method infeasible \citep[e.g.,][]{reynolds_chiu}. Despite the broken stick being the limiting case of the bent cable, reliable inference for the underlying population transition of $T_c$ requires that the stick population be an explicit component of a mixture model comprising both stick and cable populations, even if the cable population dominates in size. Moreover, the mixture allows better inference for the CTPs for separate populations (not presumed identical a priori); we have evidence that the population in the rat study consists of more than just A or just G, so that the inference (for the CTP and other parameters) would be incorrect if we did not use the mixture. Therefore,~our~extension~of bent-cable regression to model longitudinal data for multiple units provides a desirable statistical tool to characterize a special type of continuous temporal trend~---~one showing a change due to a shock that exhibits both gradual~and~abrupt~transitions. Although it would require further subject-matter research to investigate the physiological reason for certain individuals to exhibit an abrupt instead of gradual change, our flexible bent-cable approach offers an empirical solution for identifying them and making integrated inference for their CTP alongside individuals who exhibit a gradual change. Our methodology, under a general regression modelling framework, can classify observational units in the same longitudinal study as exhibiting either an abrupt or gradual transition. It provides not only inference that is more realistic, but also insights into the underlying behaviour within a population. As such, it is applicable to the rat model for induced hypothermia, and potentially to a wide variety of~other~situations. Also, if there were enough observations to support, say, a third population, our method could be easily extended to include a third component of the mixture model.

Our method is intended for only stationary AR($p$), $p\ge0$, processes for $\{\epsilon_{ij}\}$, though simulations suggested that assuming an AR(0) structure even when serial correlation exists among repeated measurements does not lead to problematic bias when characterizing the populations. In this case, serial correlation is induced by the random regression coefficients. Some directions of extension to address this and other limitations are suggested in Appendix~\ref{s:future_work}. Overall, the flexible mixture bent-cable model for~longitudinal~data~as proposed in this paper has many attractive~properties~and has allowed us to model data~from, and provide informative interpretations for, a scientific problem~of~great~practical~interest.

\section*{Appendices}
\begin{asection}
\subsection{Approximating Fitted Values}
\label{s:fitted_values}
The parameter vectors $\boldsymbol{\beta}_i$ and $\boldsymbol{\alpha}_i$ have their own posterior distributions, so the bent-cable function $f_{ij}$ itself has a posterior distribution at each observed time point $t_{ij}$, $j = 1,2,\ldots,n_i$. We consider the posterior of the bent-cable function to produce the fitted values by taking the MCMC sample means of the bent-cable function. So, the bent cable for the $i^{th}$ individual at observed time $t_{ij}$ is~$f_{ij} = \beta_{0i} + \beta_{1i} + \beta_{2i} q_{ij}$,~and the corresponding fitted~values~are
\begin{equation*}
\hat{f}_{ij}=\frac{1}{T}\sum_{s=1}^{T}{\big(\beta_{0i}^{(s)}+\beta_{1i}^{(s)}t_{ij}+\beta_{2i}^{(s)}q_{ij}^{(s)}\big)},~~ j = 1, 2, \ldots, n_i
\vspace{0.7cm}
\end{equation*}
with\vspace{-1.16cm}
\begin{equation*}\hspace{1cm}\small
q_{ij}^{(s)}=\frac{\big(t_{ij}-\tau_i^{(s)}+\gamma_i^{(s)}\big)^2}{4\gamma_i^{(s)}} \mathbf{1}\big\{|t_{ij}-\tau_i^{(s)}|\leq\gamma_i^{(s)}\big\}+ \big(t_{ij}-\tau_i^{(s)}\big)\mathbf{1}\big\{t_{ij}-\tau_i^{(s)} > \gamma_i^{(s)}\big\},
\end{equation*}
where $T$ is the length of the MCMC samples.

\subsection{Model Selection}
\label{s:model_selection}
Model selection procedure is carried out by comparing DICs. We initially consider four flexible mixture bent-cable models assuming $\epsilon_{ij}$'s to follow an AR($p$) process for $p=0,1,2,3$.  Note that we consider a conditional likelihood framework for an AR($p$) process, where the initial $p$ observations for each $i$ are treated as known. Therefore, as suggested by \citet{chiu_lockhart}, the analyses were initially performed on a reduced dataset to make the DICs comparable for $p=0, 1,2,3$. Specifically, we consider $(y_{i,4},y_{i,5},...,y_{i,n_i})'$ as the response vector (random) for the $i^{th}$ individual for all comparisons, while $(y_{i,3})$, $(y_{i,2},y_{i,3})'$ and $(y_{i,1},y_{i,2},y_{i,3})'$ are treated as known for $p=1, 2$ and $3$, respectively. That is, we dropped the first $3-p$ observations for each $i$ for $p=0,1,2,3$, respectively.

Preliminary analysis (not shown) reveals that the data exhibit nonstationarity when assuming $p>0$ for \{$\epsilon_{ij}$\}: the proportion of draws from the full conditional of~$\boldsymbol{\phi}$~for which~the stationarity condition is satisfied is close to zero. For example, an AR(1) assumption with prior $\phi \sim N(0,10^4)$ leads to $\hat{\phi}=0.99$ with $\text{DIC}\approx 5.16\times 10^{8}$. To achieve stationarity, we also consider $\phi \sim N(0,5\times10^{-5})$ and $\phi \sim N(0,2.5\times10^{-5})$ that lead to $\text{DIC}\approx 10722$ and $-14004$, respectively. In addition, the fitted coefficients change depending on the prior variance for $\phi$. Nonstationarity was also observed for AR(2) and AR(3) assumptions.

In light of the extreme sensitivity to the prior for $\boldsymbol{\phi}$ while assuming stationarity, we assume within-individual conditional independence (AR(0)), such that within-individual dependence among repeated measurements is due solely to the inclusion of the random effects $\boldsymbol{\theta}_i$'s. Our simulations~(Section~\ref{s:simulation})~reveal that though the estimates of the $\sigma_i$'s could be~less reliable, the flexible methodology can perform well with respect to the population parameters even for a misspecified correlation structure for the~$\epsilon_{ij}$'s.~Since~our~main goal is to make inference about the populations, we report results for~AR(0) with $\text{DIC}_{\text{Flexible}}\approx -14729$ (smallest observed) in Section~\ref{s:analysis}. We also analyzed the data using Models G and A for AR(0), for which $\text{DIC}_{\text{G}}\approx -14570$ and $\text{DIC}_{\text{A}}\approx -13819$, that is, our flexible mixture bent-cable model yielded better goodness of fit. Finally, note that the reported inference is actually based on the full data, i.e., using the reduced dataset as described above was solely for the purpose of DIC comparisons.

\subsection{Possible Extensions}
\label{s:future_work}
Although tailored for the rat study, the mixture longitudinal bent cable is perceivably applicable to other studies involving profiles that exhibit abrupt and/or gradual transitions of temporal trend. Thus, extensions of our framework may be desirable in some cases. For example, our framework is intended for only stationary AR($p$), $p\ge0$, processes for within-individual noise, and it might be useful to extend the framework to specifically account for nonstationarity. Other possible extensions include (i) with sufficient data, allowing for additional populations to be considered in the mixture, and (ii) allowing the variation of profiles to depend on both random and systematic components (covariates).
\end{asection}

\newpage
\titleformat{\section}{\normalfont\Large\bfseries\centering}{S\thesection.}{1em}{}
\titleformat{\subsection}{\normalfont\large\bfseries}{S\thesubsection}{1em}{}
\setlength{\baselineskip}{24pt} 
\setcounter{section}{0}

\section*{\centering SUPPLEMENTARY MATERIAL}

\section{Choice of the Hyperparameters}
\label{s:hyperparameters}
Values of the hyperparameters reflect our prior knowledge. When little is reliably known about the individual trajectories beyond its functional form of the bent-cable, it is reasonable to choose the hyperprior values that lead to fairly vague, minimally informative priors~\citep{carlin}.

The choice of a mean vector (e.g., $\mathbf{h}_1$, $\mathbf{h}_2$ or $\mathbf{h}_3$) has very little effect on Bayesian estimation, as long as the respective variance parameters (diagonal elements of $\mathbb{H}_1$, $\mathbb{H}_2$ or $\mathbb{H}_3$, respectively) are taken to be very large which lead to flat priors \citep{song}. Therefore, a common practice is to choose a zero mean vector and a covariance matrix, say, $\mathbb{H}_1$ such that $\mathbb{H}_1^{-1}\approx \mathbb{O}$, where $\mathbb{O}$ is a matrix with all its elements zero \citep{davidian}.

We use the parameterization of the gamma distribution as given in \citet{chib}. For example, $[\sigma_{i}^{-2}|d_0,d_1] \sim G(\frac{d_0}{2},\frac{d_1}{2})$. Small values of the hyperprior parameters (e.g., $d_0=d_1=10^{-4}$) lead to a diffuse prior.

We use the parameterization of the Wishart distribution as given in \citet{carlin}. For example, $[\Sigma_\beta^{-1}|\nu_1,\mathbb{A}_1] \sim W(\nu_1,(\nu_1 \mathbb{A}_1)^{-1})$. Setting the degrees of freedom equal to the order of the scale matrix (e.g. 3 for the prior of $\Sigma_\beta^{-1}$) makes a Wishart prior nearly flat \citep{wake94}. The matrix $\mathbb{A}_1$ (or $\mathbb{A}_2$) is chosen to be an approximate prior estimate of $\Sigma_\beta$ (or $\Sigma_\alpha$). In the absence of such prior knowledge, one may use the sample covariance matrix of the individual-specific estimates of the regression coefficients; the {\sf R} \citep{r} library ``bentcableAR'' \citep{chiuR} for single profile bent-cable regression can be useful in this regard.

Since $0<\omega<1$, we choose the beta distribution $[\omega|c_0,c_1] \sim B(c_0,c_1)$ in our model. In the absence of prior information, one may choose $c_0=c_1=1$ which leads to $U(0,1)$ distribution.

\section{Full Conditionals}
\label{s:full_conditional}
For the full conditionals of the model parameters, let
\begin{itemize}\onehalfspacing
\item $z_{ij}=y_{ij}-\sum_{k=1}^p{\phi_k~y_{i,j-k}}$, ~$x_{ij}=t_{ij}-\sum_{k=1}^p{\phi_k~t_{i,j-k}}$,~ $r_{ij}=q_{ij}-\sum_{k=1}^p{\phi_k~q_{i,j-k}}$, $\mathbf{z}_i=(z_{i,p+1},\ldots,z_{i,n_i})'$,~ $\mathbf{x}_i=(x_{i,p+1},\ldots,zx_{i,n_i})'$,~ $\mathbf{r}_i=(r_{i,p+1},\ldots,r_{i,n_i})'$ and $\mathbb{X}_i=(1-\sum_{k=1}^p{\phi_k},~\mathbf{x}_i,~\mathbf{r}_i)$;\vspace{0.05cm}
\item $\epsilon_{ij}=y_{ij}-\beta_{0i}-\beta_{1i}~t_{ij}-\beta_{2i}~q_{ij}$~ for $j=p+1,\ldots,n_i$, \vspace{0.05cm} ~$\boldsymbol{\epsilon}_i=(\epsilon_{i,p+1},\ldots,\epsilon_{i,n_i})'$~ and $\mathbb{V}^{-1}=\sum_{i=1}^{m}{\sigma_{i}^{-2}~\mathbb{W}_i'~\mathbb{W}_i}+
    \mathbb{H}_3^{-1}$, where $\mathbb{W}_i$\; is a $(n_i-p) \times p$ matrix with the $k^{th}$ row given by $(\epsilon_{i,k+p-1}, \epsilon_{i,k+p-2},\ldots, \epsilon_{i,k})$;\vspace{0.05cm}
\item $m_A=\sum_{i=1}^{m}{(1-I_i)}$ and $m_G=\sum_{i=1}^{m}{I_i}$;\vspace{0.05cm}
\item $\boldsymbol{\xi}_i=\log{\boldsymbol{\alpha}_i}=(\log{\gamma}_i,\log{\tau}_i)'$ and $\kappa_i=\log{\tau}_i$;\vspace{0.05cm}
\item $\tilde{\boldsymbol{\beta}}=\sum_{i=1}^{m}{\boldsymbol{\beta}_i}$,~ $\tilde{\boldsymbol{\xi}}=\sum_{i=1}^{m}{I_i~ \boldsymbol{\xi}_i}$,~ and $\tilde{\kappa}=\sum_{i=1}^{m}{(1-I_i)~\kappa_i}$;\vspace{0.05cm}
\item $\mathbb{M}_i^{-1}=\sigma_{i}^{-2}~\mathbb{X}_i'~\mathbb{X}_i+\Sigma_\beta^{-1}$;\vspace{0.05cm}
\item $\mathbb{U}_1^{-1}=m~\Sigma_\beta^{-1}+\mathbb{H}_1^{-1}$ and $\mathbb{U}_2^{-1}=m_G~\Sigma_\alpha^{-1}+\mathbb{H}_2^{-1}$.
\end{itemize}

An appealing feature of the bent-cable function is that it is partially linear -- given $\boldsymbol{\alpha}_i$, $f(t_{ij},\boldsymbol{\theta}_i)$ is linear -- and we can exploit this fact to derive a closed-form full conditional for $\boldsymbol{\beta}_i$. However, the full conditional of $\boldsymbol{\alpha}_i$ can be expressed only up to a proportionality constant, and is given by\vspace{0.1cm}
\begin{equation*}
\begin{split}
\pi(\boldsymbol{\alpha}_i|.) &\propto \exp{\Big\{-\frac{1}{2\sigma^2_{i}}(\mathbf{z}_i-\mathbb{X}_i~\boldsymbol{\beta}_i)'(\mathbf{z}_i-
\mathbb{X}_i~\boldsymbol{\beta}_i)\Big\}}
\times \Big[\frac{1}{\tau_i}\exp{\Big\{-\frac{1}{2\sigma_{\tau_A}^2}(\kappa_i-\mu_{\tau_A})^2\Big\}} \Big]^{1-I_i} \\
& \times \Big[\frac{1}{\gamma_i \tau_i}\exp{\Big\{-\frac{1}{2}(\boldsymbol{\xi}_i-\boldsymbol{\mu}_\alpha)'~\Sigma_\alpha^{-1}~(\boldsymbol{\xi}_i-
\boldsymbol{\mu}_\alpha)\Big\}}\Big]^{I_i}.
\end{split}
\end{equation*}\vspace{-0.2cm}

\noindent The full conditionals of the remaining parameters are
\begin{equation*}
[\boldsymbol{\beta}_i|.] \sim
N_3\Big(\mathbb{M}_i~ \big(\sigma_{i}^{-2}~\mathbb{X}_i'~\mathbf{z}_i
+\Sigma_\beta^{-1}~\boldsymbol{\mu}_\beta\big),~\mathbb{M}_i\Big),
\label{ch4.e:full_conditional_beta_i}\vspace{-0.1cm}
\end{equation*}
\begin{equation*}
[\boldsymbol{\mu}_\beta|.] \sim
N_3\Big(\mathbb{U}_1~\big(\Sigma_\beta^{-1}~\tilde{\boldsymbol{\beta}}+ \mathbb{H}_1^{-1}~\mathbf{h}_1\big),~\mathbb{U}_1\Big), \label{ch4.e:full_conditional_mu_beta}
\end{equation*}
\begin{equation*}
[\boldsymbol{\mu}_\alpha|.] \sim
N_2\Big(\mathbb{U}_2~\big(\Sigma_\alpha^{-1}~\tilde{\boldsymbol{\xi}}+ \mathbb{H}_2^{-1}~\mathbf{h}_2\big),~\mathbb{U}_2\Big), \label{ch4.e:full_conditional_mu_alpha}\vspace{0.1cm}
\end{equation*}
\begin{equation*}
[\mu_{\tau_A}|.] \sim
N\bigg(\frac{\sigma_{\tau_A}^{-2}~\tilde{\kappa}+a_1^{-1}~a_0}{m_A~\sigma_{\tau_A}^{-2}+a_1^{-1}},~ \frac{1}{m_A~\sigma_{\tau_A}^{-2}+a_1^{-1}}\bigg),
\label{ch4.e:full_conditional_mu_tau_A}\vspace{0.1cm}
\end{equation*}
\begin{equation*}
[\Sigma_\beta^{-1}|.] \sim
W\bigg(m+\nu_1,~\Big[\sum_{i=1}^{m} {(\boldsymbol{\beta}_i-\boldsymbol{\mu}_\beta)~(\boldsymbol{\beta}_i-\boldsymbol{\mu}_\beta)'} +\nu_1\mathbb{A}_1\Big]^{-1}\bigg),
\label{ch4.e:full_conditional_Sigma_beta}\vspace{0.1cm}
\end{equation*}
\begin{equation*}
[\Sigma_\alpha^{-1}|.] \sim
W\bigg(m_G+\nu_2,~\Big[\sum_{i=1}^{m} {I_i(\boldsymbol{\xi}_i-\boldsymbol{\mu}_\alpha)~(\boldsymbol{\xi}_i-\boldsymbol{\mu}_\alpha)'} +\nu_2\mathbb{A}_2\Big]^{-1}\bigg),
\label{ch4.e:full_conditional_Sigma_alpha}\vspace{0.1cm}
\end{equation*}
\begin{equation*}
[\sigma_{\tau_A}^{-2}|.] \sim
G\bigg(\frac{m_A+b_0}{2},\frac{\sum_{i=1}^{m}{(1-I_i)(\kappa_i-\mu_{\tau_A})^2}+b_1}{2}\bigg),
\label{ch4.e:full_conditional_sigma_tau_A}\vspace{0.1cm}
\end{equation*}
\begin{equation*}
[\sigma_{i}^{-2}|.] \sim
G\bigg(\frac{n_i-p+d_0}{2},~\frac{(\mathbf{z}_i-\mathbb{X}_i~\boldsymbol{\beta}_i\big)'
(\mathbf{z}_i-\mathbb{X}_i~\boldsymbol{\beta}_i\big)+d_1}{2}\bigg),
\label{ch4.e:full_conditional_sigma_i}\vspace{0.1cm}
\end{equation*}
\begin{equation*}
[\boldsymbol{\phi}|.]
\sim N_p\bigg(\mathbb{V}\Big(\sum_{i=1}^{m}{\sigma_{i}^{-2}~\mathbb{W}_i'~\epsilon_i}+
\mathbb{H}_3^{-1}~\mathbf{h}_3\Big),~\mathbb{V}\bigg),
\label{ch4.e:full_conditional_phi}\vspace{0.1cm}
\end{equation*}
\begin{equation*}
[\omega|.]
\sim B(m_G+c_0,m_A+c_1).
\label{ch4.e:full_conditional_omega}
\end{equation*}

\section{Fitted Population Curves for the Rat Profiles}
\label{s:curves}
\begin{figure}[ht!]\vspace{0.3cm}
\centerline{\includegraphics[width=4in,height=3.4in]{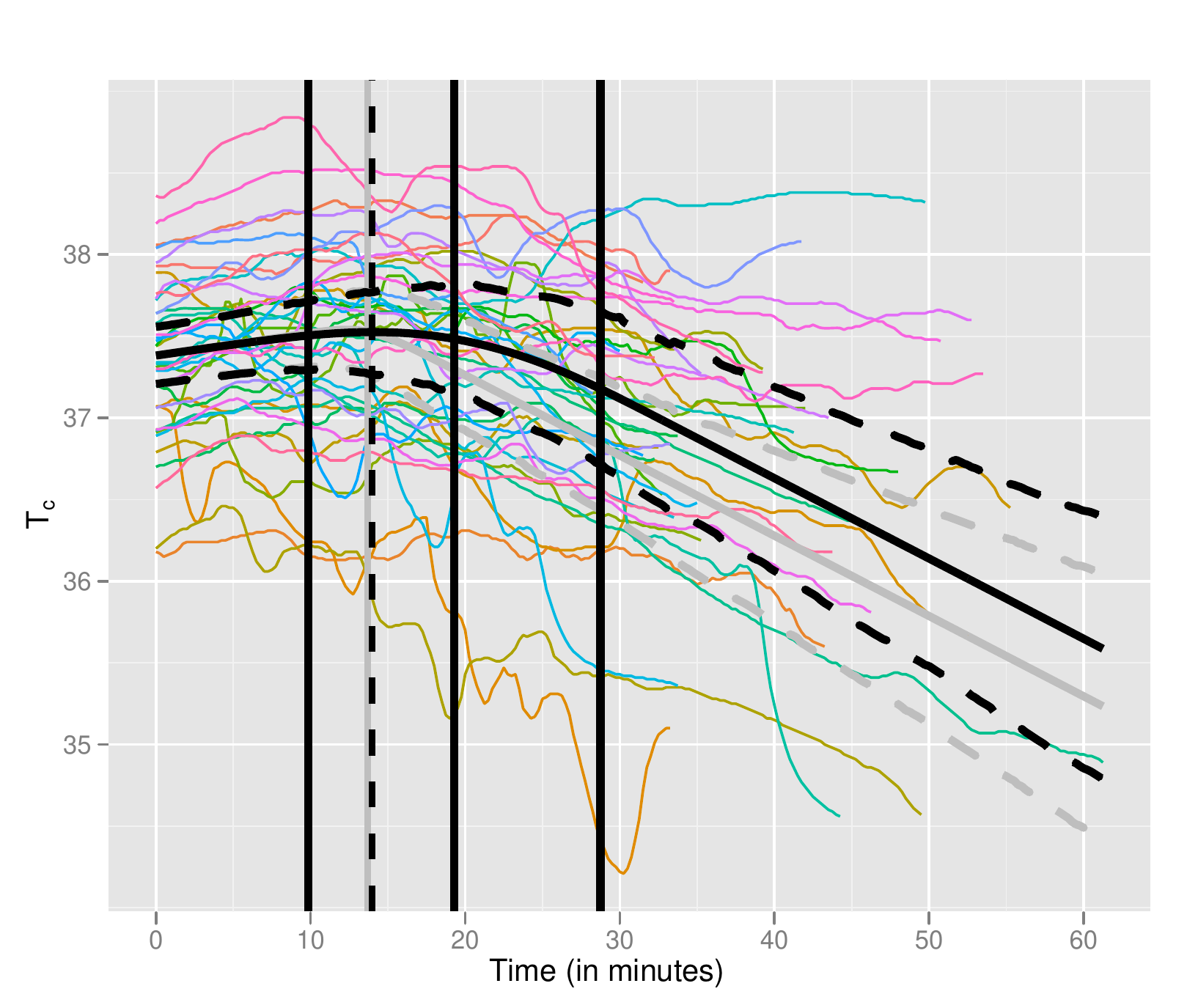}}
\caption{Observed profiles and fitted population curves (solid, in bold) with 95\% pointwise credible intervals (dotted curves) for the two populations (grey for A and black for G). The model fit is produced assuming conditional within-individual independence. The estimated transition for G \big(i.e. $\widehat{\mbox{\fontsize{7}{9}\selectfont $\mathcal{M}_\tau\pm\mathcal{M}_\gamma$}}$\big) is marked by solid black vertical lines, and that for A \big(i.e. $\mbox{\fontsize{7}{9}\selectfont $\widehat{\mathcal{M}}_{\tau_A}$}$\big) by the grey vertical line. The estimated CTP for G is indicated by the dotted vertical line.}
\label{f:ratpopfit1}\vspace{0.3cm}
\end{figure}

In the main text, we presented the rat data analysis; the fitted population curves were displayed in Figure~3. Here, in Figure~\ref{f:ratpopfit1}, we reproduce the same figure but overlaid with all 38 rat profiles; the population fitted curves are displayed in bold. Figure~\ref{f:ratpopfit1} displays the whole range of shapes of the rat profiles. It also shows that the profiles are well represented by the population fitted curves.

\section{Detailed Simulation Results}
\label{s:simresults}
Main findings of our simulations were summarized in the main text. Here we present the numerical results with supplementary information.

\subsection{Scenario 1}
\label{ass:scenario1}
Simulation results for Scenario 1 are presented in Tables~\ref{t:pop1}~-~\ref{t:innovation1}\hspace{0.05cm}. In the main text, we discussed our findings for the population regression coefficients of Table~\ref{t:pop1}\vspace{0.3cm}.
\begin{table}[ht!]\centering
\caption{Simulation scenario 1 results with $n_i=150$ for all $i$ and $m=20$: average of 500 posterior means of the population regression coefficients and the AR parameters; also coverage of 95\% credible intervals.} \footnotesize
\begin{threeparttable}
  \begin{tabular}{l r cc c}
  \hline
    & & \multicolumn{1}{c}{Simulated $\omega=0.90$} & & \multicolumn{1}{c}{Simulated $\omega=0.95$}\\
    \cline{3-3} \cline{5-5}
    & & \multicolumn{1}{c}{Analysis using Model G} & & \multicolumn{1}{c}{Analysis using Model G}\\
    \cline{3-3} \cline{5-5}
    &True & Mean, Coverage& & Mean, Coverage  \\
  \hline
  $\mu_0$ &$244.00$ &$244.33,0.95$  &  & $244.51, 0.96$\\
  $\mu_1$ &$0.50$ &$0.49,0.93$  &  & $0.49,0.94$\\
  $\mu_2$ &$-0.75$ & $-0.78,0.91$ &  & $-0.78,0.92$\\
  $\mu_\gamma$ &$3.00$ &$2.88,0.86$  &  & $2.94,0.92$\\
  $\mu_\tau$ &$4.00$ &$4.04,0.92$  &  & $4.02,0.95$\\
  $\mu_{\tau_A}$ &$4.50$ &$-$  &  & $-$\\
  $\phi$ &$0.70$ &$0.71,0.93$  &  & $0.71,0.93$\\
  \hline
\end{tabular}
\end{threeparttable}
\label{t:pop1}\vspace{0.1cm}
\end{table}

\begin{table}[ht!]\centering
\caption{Simulation scenario 1 results with $n_i=150$ for all $i$ and $m=20$: average of 500 posterior means (medians for the variance parameters) of the variances and covariances ($\Sigma_\beta$ and $\Sigma_\alpha$) in the priors for the random regression coefficients; also coverage of 95\% credible intervals.} \footnotesize \vspace{0.05cm}
\begin{threeparttable}
  \begin{tabular}{l r cc c}
  \hline
    & & \multicolumn{1}{c}{Simulated $\omega=0.90$} & & \multicolumn{1}{c}{Simulated $\omega=0.95$}\\
    \cline{3-3} \cline{5-5}
    & & \multicolumn{1}{c}{Analysis using Model G} & & \multicolumn{1}{c}{Analysis using Model G}\\
    \cline{3-3} \cline{5-5}
    &True & Mean, Coverage& & Mean, Coverage  \\
  \hline
  $(\Sigma_\beta)_{11}$ &$125.00$ &$123.65,0.99$  &  & $123.38,0.98$\\
  $(\Sigma_\beta)_{22}$ &$0.03$ &$0.03,0.98$  &  & $0.03,0.97$\\
  $(\Sigma_\beta)_{33}$ &$0.03$ & $0.03,0.99$ &  & $0.03,0.99$\\
  $(\Sigma_\beta)_{12}$ &$-1.00$ &$-0.95,0.96$  &  & $-0.94,0.97$\\
  $(\Sigma_\beta)_{13}$ &$0.50$ &$0.57,0.99$  &  & $0.57,0.99$\\
  $(\Sigma_\beta)_{23}$ &$-0.01$ &$-0.01,0.99$  &  & $-0.01,0.99$\\
  $(\Sigma_\alpha)_{11}$ &$0.020$ &$0.114,0.69$  &  & $0.067,0.82$\\
  $(\Sigma_\alpha)_{22}$ &$0.030$ &$0.054,0.62$  &  & $0.043,0.78$\\
  $(\Sigma_\alpha)_{12}$ &$0.005$ &$-0.051,0.68$  &  & $-0.024,0.83$\\
  $\sigma_{\tau_A}^2$ &$0.050$ &$-$  &  & $-$\vspace{0.05cm}\\
  \hline
\end{tabular}
\end{threeparttable}
\label{t:cov1}
\end{table}

\begin{table}[ht!]\centering
\caption{Simulation scenario 1 results with $n_i=150$ for all $i$ and $m=20$: average of 500 posterior medians of the innovation variances; also coverage of 95\% credible intervals.} \footnotesize
\begin{threeparttable}
  \begin{tabular}{l r cc c}
  \hline
    & & \multicolumn{1}{c}{Simulated $\omega=0.90$} & & \multicolumn{1}{c}{Simulated $\omega=0.95$}\\
    \cline{3-3} \cline{5-5}
    & & \multicolumn{1}{c}{Analysis using Model G} & & \multicolumn{1}{c}{Analysis using Model G}\\
    \cline{3-3} \cline{5-5}
    &True & Mean, Coverage& & Mean, Coverage  \\
  \hline
  $\sigma_{1}^2$ &$0.34$ &$0.35, 0.97$  &  & $0.35, 0.94$\\
  $\sigma_{2}^2$ &$1.12$ &$1.14, 0.95$  &  & $1.12, 0.94$\\
  $\sigma_{3}^2$ &$1.75$ & $1.78, 0.95$ &  & $1.76, 0.96$\\
  $\sigma_{4}^2$ &$0.42$ &$0.42, 0.95$  &  & $0.42, 0.96$\\
  $\sigma_{5}^2$ &$0.74$ &$0.76, 0.94$  &  & $0.74, 0.94$\\
  $\sigma_{6}^2$ &$2.06$ &$2.08, 0.95$  &  & $2.08, 0.95$\\
  $\sigma_{7}^2$ &$1.16$ &$1.16, 0.94$  &  & $1.16, 0.93$\\
  $\sigma_{8}^2$ &$1.28$ &$1.29, 0.93$  &  & $1.27, 0.93$\\
  $\sigma_{9}^2$ &$0.16$ &$0.16, 0.95$  &  & $0.16, 0.96$\\
  $\sigma_{10}^2$ &$0.77$ &$0.78, 0.96$  &  & $0.77, 0.94$\\
  $\sigma_{11}^2$ &$0.04$ &$0.04, 0.96$  &  & $0.04, 0.95$\\
  $\sigma_{12}^2$ &$0.03$ &$0.03, 0.96$  &  & $0.03, 0.94$\\
  $\sigma_{13}^2$ &$0.91$ &$0.92, 0.96$  &  & $0.92, 0.95$\\
  $\sigma_{14}^2$ &$1.96$ &$1.97, 0.94$  &  & $1.96, 0.95$\\
  $\sigma_{15}^2$ &$0.32$ &$0.33, 0.96$  &  & $0.32, 0.96$\\
  $\sigma_{16}^2$ &$2.02$ &$2.02, 0.95$  &  & $2.05, 0.95$\\
  $\sigma_{17}^2$ &$0.89$ &$0.90, 0.95$  &  & $0.90, 0.96$\\
  $\sigma_{18}^2$ &$0.90$ &$0.90, 0.94$  &  & $0.91, 0.95$\\
  $\sigma_{19}^2$ &$0.82$ &$0.83, 0.95$  &  & $0.83, 0.95$\\
  $\sigma_{20}^2$ &$2.89$ &$2.93, 0.97$  &  & $2.91, 0.96$\\
  \hline
\end{tabular}
\end{threeparttable}
\label{t:innovation1}
\end{table}

We calculate the percentage closer to the true value for $\mu_\gamma$ (given on page 16 of the main text) as follows. The true $\mu_\gamma$ is $3.00$, whereas the averages of the posterior means are $2.88$ and $2.94$ for $\omega=0.90$ and $0.95$, respectively. Then, the average of the posterior means when $\omega=0.95$ is $100\{(3.00-2.88) - (3.00-2.94)\}/(3.00-2.88)=50\%$ closer to the true value when compared $\omega=0.90$. Similarly, the coverage for $\mu_\gamma$ is $100\{(0.95-0.86) - (0.95-0.92)\}/(0.95-0.86)\approx 67\%$ closer to the nominal $0.95$.

From Table~\ref{t:cov1},~we see that coverage~probabilities~for the elements of $\Sigma_\beta$ are all close~to $0.99$. Since $\Sigma_\alpha$ takes into account both~A~and~G,~large~variabilities among the~$\gamma_i$'s and~$\tau_i$'s are expected. In our simulation study, large variabilities are indeed reflected through the overestimation for each of the variance parameters $(\Sigma_\alpha)_{11}$ and $(\Sigma_\alpha)_{22}$. Because of the overestimation, we also see low coverage~probabilities.~Moreover, for a true~$(\Sigma_\alpha)_{11}$~of~$0.020$, the average of its posterior medians is $0.114$ when $\omega=0.90$, and is $0.067$ when $\omega=0.95$. Hence, estimation of $(\Sigma_\alpha)_{11}$ is more accurate for $\omega$ close to 1; the same is true for $(\Sigma_\alpha)_{22}$.

Finally, the average of the posterior medians for each $\sigma_i^2$ is close to the truth, and coverage probabilities are all close the the nominal $0.95$; see Table~\ref{t:innovation1}. That is, misspecifying the model as Model G, when in reality both populations A and G exist with G being dominant, has virtually no effects on the estimates of $\sigma_i^2$'s.\vspace{0.25cm}

\subsection{Scenarios 2 and 3}
\label{ass:scenario23}
Simulation results for Scenarios 2 and 3 are presented in Tables~\ref{t:pop2}~-~\ref{t:innovation2}. We discussed our findings for the population regression coefficients (Table~\ref{t:pop2}) in the main text.\vspace{0.3cm}

{\renewcommand{\arraystretch}{1.3}
\renewcommand{\tabcolsep}{0.12cm}
\begin{table}[ht!]\centering
\caption{Simulation scenarios 2 and 3 results with $n_i=150$ for all $i$ and $m=20$: average of 500 posterior means of the mixing proportion $\omega$, population regression coefficients and the AR parameters; also coverage of 95\% credible intervals.} \footnotesize
\begin{threeparttable}
  \begin{tabular}{l r cccc ccc}
  \hline
    & & \multicolumn{3}{c}{Simulated $\epsilon_{ij}$'s: AR(1)} & & \multicolumn{3}{c}{Simulated $\epsilon_{ij}$'s: AR(2)}\\
    \cline{3-5} \cline{7-9}
    & & \multicolumn{3}{c}{Analysis assuming} & & \multicolumn{3}{c}{Analysis assuming}\\
    & & AR(2) & AR(1) & AR(0) & & AR(2) & AR(1) & AR(0)\\
    \cline{3-5} \cline{7-9}
    & & Mean, & Mean, & Mean, & & Mean, & Mean, & Mean, \\
    & True& Coverage & Coverage&Coverage& &Coverage &Coverage &Coverage\\
  \hline
  $\omega$ &$0.50$ & $0.52,0.98$ & $0.52,0.99$ & $0.52,0.97$  & & $0.52,0.97$ & $0.51,0.95$ & $0.52,0.96$ \\
  $\mu_0$ &$244.00$ &$243.94,0.97$ & $244.45,0.94$ & $244.37,0.96$& & $244.64,0.96$ &  $244.30,0.95$ & $244.45,0.96$ \\
  $\mu_1$ & $0.50$ &$0.50,0.97$ & $0.48,0.94$ &  $0.49,0.95$ & & $0.48,0.94$ & $0.48,0.95$ & $0.48,0.93$\\
  $\mu_2$ & $-0.75$ &$-0.75,0.95$ & $-0.77,0.92$ & $-0.78,0.92$ & & $-0.77,0.95$ & $-0.78,0.93$ & $-0.77,0.92$\\
  $\mu_\gamma$ & $3.00$ &$2.95,0.95$ &$2.94,0.95$ & $2.95,0.93$  & & $2.97,0.96$ & $2.95,0.97$ & $2.96,0.93$\\
  $\mu_\tau$ & $4.00$ &$4.02,0.98$ &$4.02,0.97$ & $4.04,0.96$ & & $4.01,0.99$ & $4.02,0.96$ & $4.04,0.92$\\
  $\mu_{\tau_A}$ & $4.50$ &$4.50,0.94$ & $4.50,0.96$ & $4.47,0.94$ & &$4.49,0.95$ & $4.49,0.97$ & $4.47,0.94$\\
  AR(1) $\phi$ & $0.70$ &$-$ & $0.71,0.93$ & $-$ & &$-$ & $0.74, -$ & $-$\\
  AR(2) $\phi_{1}$ & $0.80$ &$0.70, 0.94$ & $-$ & $-$ & &$0.80,0.95$ & $-$ & $-$\\
  AR(2) $\phi_{2}$& $-0.10$ &$0.005, 0.92$ & $-$ & $-$& &$-0.10,0.95$ & $-$ & $-$ \\
  \hline
\end{tabular}
\end{threeparttable}
\label{t:pop2}\vspace{0cm}
\end{table}

\begin{table}[ht!]\centering \vspace{-0.32cm}
\caption{Simulation scenarios 2 and 3 results with $n_i=150$ for all $i$ and $m=20$: average of 500 posterior means (medians for the variance parameters) of the variances and covariances ($\sigma_{\tau_A}^2$, $\Sigma_\beta$ and $\Sigma_\alpha$) in the priors for the random regression coefficients; also coverage of 95\% credible intervals.} \footnotesize
\begin{threeparttable}
  \begin{tabular}{l r cccc ccc}
  \hline
    & & \multicolumn{3}{c}{Simulated $\epsilon_{ij}$'s: AR(1)} & & \multicolumn{3}{c}{Simulated $\epsilon_{ij}$'s: AR(2)}\\
    \cline{3-5} \cline{7-9}
    & & \multicolumn{3}{c}{Analysis assuming} & & \multicolumn{3}{c}{Analysis assuming}\\
    & & AR(2) & AR(1) & AR(0) & & AR(2) & AR(1) & AR(0)\\
    \cline{3-5} \cline{7-9}
    & & Mean, & Mean, & Mean, & & Mean, & Mean, & Mean, \\
    & True& Coverage & Coverage&Coverage& &Coverage &Coverage &Coverage\\
  \hline
  $(\Sigma_\beta)_{11}$ & $125.00$ &$126.13,0.98$ &$126.78,0.97$ & $124.32,0.98$ & & $124.39,0.97$ & $123.11,0.98$ & $125.52,0.98$\\
  $(\Sigma_\beta)_{22}$ & $0.03$ &$0.03,0.99$ &$0.03,0.99$ & $0.03,0.98$& &  $0.03,0.97$ & $0.03,0.98$ & $0.03,0.98$ \\
  $(\Sigma_\beta)_{33}$ & $0.03$ &$0.03,0.98$ & $0.03,0.99$ & $0.03,0.97$& & $0.03,0.98$ & $0.03,0.97$ & $0.03,0.99$ \\
  $(\Sigma_\beta)_{12}$ & $-1.00$ &$-1.05,0.98$ & $-0.97,0.95$ & $-0.98,0.97$& & $-0.97,0.96$ & $-0.93,0.97$ & $-0.97,0.97$ \\
  $(\Sigma_\beta)_{13}$ & $0.50$ &$0.49,0.99$ &$0.60,0.98$ & $0.53,0.99$ & &  $0.54,0.98$ & $0.53,0.99$ & $0.49,0.99$\\
  $(\Sigma_\beta)_{23}$ & $-0.01$ &$-0.01,0.97$ &$-0.01,0.98$ & $-0.01,0.98$ & &  $-0.01,0.98$ & $-0.01,0.99$ & $-0.01,0.98$\\
  $(\Sigma_\alpha)_{11}$ & $0.020$ &$0.021,1.00$ &$0.020,0.99$ & $0.059,0.78$ & & $0.021,0.99$ & $0.036,1.00$ & $0.058,0.80$\\
  $(\Sigma_\alpha)_{22}$ & $0.030$ & $0.031,0.99$ & $0.031,0.99$ & $0.045,0.91$ & & $0.032,0.99$ & $0.032,0.99$ & $0.043,0.93$\\
  $(\Sigma_\alpha)_{12}$ & $0.005$ &$0.0002,1.00$ & $0.001,1.00$ & $-0.007,0.95$ & &$0.001,0.99$ & $0.0003,1.00$ & $-0.006,0.95$\\
  $\sigma_{\tau_A}^2$ & $0.050$ &$0.57,0.97$ & $0.059,0.98$ & $0.069,0.98$& &$0.059,0.95$ & $0.059,0.97$ & $0.073,0.97$ \vspace{0.02cm}\\
  \hline
\end{tabular}
\end{threeparttable}
\label{t:cov2}
\end{table}

\begin{table}[ht!]\centering
\caption{Simulation scenarios 2 and 3 results with $n_i=150$ for all $i$ and $m=20$: average of 500 posterior medians of the innovation variances; also coverage of 95\% credible intervals.} \footnotesize \vspace{0.05cm}
\begin{threeparttable}
  \begin{tabular}{l r cccc ccc}
  \hline
    & & \multicolumn{3}{c}{Simulated $\epsilon_{ij}$'s: AR(1)} & & \multicolumn{3}{c}{Simulated $\epsilon_{ij}$'s: AR(2)}\\
    \cline{3-5} \cline{7-9}
    & & \multicolumn{3}{c}{Analysis assuming} & & \multicolumn{3}{c}{Analysis assuming}\\
    & & AR(2) & AR(1) & AR(0) & & AR(2) & AR(1) & AR(0)\\
    \cline{3-5} \cline{7-9}
    & & Mean, & Mean, & Mean, & & Mean, & Mean, & Mean, \\
    & True& Coverage & Coverage&Coverage& &Coverage &Coverage &Coverage\\
  \hline
  $\sigma_{1}^2$ & $0.34$ &$0.35, 0.96$ & $0.35, 0.96$ & $0.58, 0.11$ & &$0.35, 0.96$ & $0.36, 0.95$ & $0.65, 0.02$\\
  $\sigma_{2}^2$ & $ 1.12$ &$1.12, 0.94$ & $1.13, 0.97 $ & $1.89, 0.10$ & &$1.12, 0.94$ & $1.16, 0.95$ & $2.09, 0.02$\\
  $\sigma_{3}^2$ & $ 1.75$ &$1.77, 0.95$ & $1.76, 0.95$ & $2.95, 0.09$& &$1.76, 0.94$ & $1.80, 0.94$ & $3.27, 0.03$ \\
  $\sigma_{4}^2$ & $ 0.42$ &$0.41, 0.94$ & $0.42, 0.95$ & $0.71, 0.08$& &$0.42, 0.96$ & $0.43, 0.92$ & $0.78, 0.02$ \\
  $\sigma_{5}^2$ & $ 0.74$ &$0.75, 0.95$ & $0.74, 0.95$ & $1.26, 0.08$ & &$0.74, 0.94$ & $0.76, 0.96$ & $1.43, 0.02$\\
  $\sigma_{6}^2$ & $ 2.06$ &$2.08, 0.96$ & $2.08, 0.94$ & $3.52, 0.09$ & &$2.09, 0.95$ & $2.10, 0.94$ & $3.86, 0.03$\\
  $\sigma_{7}^2$ & $ 1.16$ &$1.16, 0.94$ &$1.16, 0.95$ & $1.95, 0.09$& & $1.16, 0.94$ & $1.18, 0.94$ & $2.19, 0.03$ \\
  $\sigma_{8}^2$ & $ 1.28$ &$1.29, 0.95$ &$1.30, 0.96$ & $2.17, 0.08$ & & $1.28, 0.95$ & $1.30, 0.96$ & $2.43, 0.03$\\
  $\sigma_{9}^2$ & $ 0.16$ &$0.16, 0.94$ & $0.16, 0.94$ & $0.27, 0.08$ & &$0.16, 0.93$ & $0.16, 0.95$ & $0.30, 0.02$\\
  $\sigma_{10}^2$ & $ 0.77$ &$0.78, 0.97$ & $0.79, 0.95$ & $1.32, 0.08$ & &$0.78, 0.95$ & $0.79, 0.95$ & $1.44, 0.03$\\
  $\sigma_{11}^2$ & $ 0.04$ &$0.04, 0.94$ & $0.04, 0.95$ & $0.06, 0.07$ & &$0.04, 0.95$ & $0.04, 0.94$ & $0.07, 0.03$\\
  $\sigma_{12}^2$ & $ 0.03$ &$0.03, 0.95$ &$0.03, 0.96$ & $0.06, 0.09$ & & $0.03, 0.94$ & $0.03, 0.95$ & $0.07, 0.02$\\
  $\sigma_{13}^2$ & $ 0.91$ &$0.92, 0.96$ &$0.92, 0.95$ & $1.55, 0.09$& & $0.91, 0.95$ & $0.93, 0.95$ & $1.70, 0.04$ \\
  $\sigma_{14}^2$ & $ 1.96$ &$1.99, 0.94$ & $1.95, 0.95$ & $3.36, 0.06$& &$1.96, 0.96$ & $1.99, 0.94$ & $3.62, 0.04$ \\
  $\sigma_{15}^2$ & $ 0.32$ &$0.32, 0.96$ & $0.33, 0.95$ & $0.55, 0.07$& &$0.33, 0.95$ & $0.33, 0.94$ & $0.61, 0.03$ \\
  $\sigma_{16}^2$ & $ 2.02$ &$2.03, 0.97$ & $2.03, 0.94$ & $3.40, 0.07$ & &$2.04, 0.96$ & $2.05, 0.95$ & $3.85, 0.03$\\
  $\sigma_{17}^2$ & $0.89 $ &$0.90, 0.94$ & $0.89, 0.94$ & $1.52, 0.09$& &$0.89, 0.95$ & $0.91, 0.94$ & $1.68, 0.04$ \\
  $\sigma_{18}^2$ & $0.90 $ &$0.90, 0.94$ & $0.90, 0.94$ & $1.53, 0.09$& &$0.91, 0.96$ & $0.92, 0.96$ & $1.70, 0.04$ \\
  $\sigma_{19}^2$ & $0.82 $ &$0.83, 0.93$ & $0.84, 0.95$ & $1.41, 0.07$& &$0.82, 0.96$ & $0.84, 0.94$ & $1.54, 0.04$ \\
  $\sigma_{20}^2$ & $2.89$ &$2.92, 0.93$ & $2.92, 0.94$ & $4.86, 0.10$& &$2.93, 0.96$ & $2.99, 0.95$ & $5.47, 0.03$ \\
  \hline
\end{tabular}
\end{threeparttable}
\label{t:innovation2}
\end{table}

For $\Sigma_\beta$, $\Sigma_\alpha$ and $\sigma^2_{\tau_A}$, coverage probabilities are all close to $0.99$ (slight over coverage) for correctly specified models (Table~\ref{t:cov2}). However, we observe poor coverage for $(\Sigma_\alpha)_{11}$ for data sets generated from AR(1) and AR(2), but using an AR(0) fit: coverage probabilities are $0.78$ and $0.80$, respectively. For such model misspecification, we also see a slight under coverage for $(\Sigma_\alpha)_{22}$ and overestimation of $(\Sigma_\alpha)_{11}$, $(\Sigma_\alpha)_{22}$ and $\sigma^2_{\tau_A}$. This suggests that underspecifying $p$ as zero may result in overestimation of transition parameter prior variances. Over-coverage, as we have observed for some parameters in all three scenarios, is of much less concern in practice than under-coverage.

Finally, we see noticeable differences in the estimates (average of the posterior medians) of $\sigma^2_{i}$'s for different $p$'s. In general, an underspecified $p$ leads to overestimation of $\sigma_{i}^2$. We observe very poor coverage (from 0.02 to 0.04) for $\sigma_i^2$ if we incorrectly analyze a data set by an AR(0) assumption when, in reality, it exhibits serial correlation over time. However, the problem is much less severe for an underspecified $p$ that is positive. Although such poor coverage may not be ideal in certain cases, of primary practical concern in the rat analysis is the inference for the population regression coefficients, for which our methodology demonstrates robustness.

\end{document}